\documentclass[nofootinbib,reprint,amsmath,amssymb,aps]{revtex4-1}
\usepackage{graphicx}
\usepackage{dcolumn}
\usepackage{bm}
\usepackage{hyperref}
\newcommand{\qm}[1]{``#1''}
\newcommand\ChangeRT[1]{\noalign{\hrule height #1}}

\begin{document}

\preprint{APS/123-QED}

\title[The three-dimensional general relativistic Poynting-Robertson effect I]{The three-dimensional general relativistic Poynting-Robertson effect I: radial radiation field}

\author{Vittorio De Falco$^{1,2,4}$} \email{vittorio.defalco@physics.cz}
\author{Pavel Bakala$^{1,5}$}\email{pavel.bakala@fpf.slu.cz}
\author{Emmanuele Battista$^{6,7}$}\email{emmanuelebattista@gmail.com}
\author{Debora Lan\v{c}ov\'a$^{1}$}
\author{Maurizio Falanga$^{2,3,4}$} 
\author{Luigi Stella$^{8}$\vspace{0.5cm}}

\affiliation{$^1$ Research Centre for Computational Physics and Data Processing, Faculty of Philosophy \& Science, Silesian University in Opava, Bezru\v{c}ovo n\'am.~13, CZ-746\,01 Opava, Czech Republic\\
$^2$ International Space Science Institute, Hallerstrasse 6, 3012 Bern, Switzerland\\
$^3$ International Space Science Institute Beijing, No.1 Nanertiao, Zhongguancun, Haidian District, 100190 Beijing,
China\\
$^4$ Departement Physik, Universit\"at Basel, Klingelbergstrasse 82, 4056 Basel, Switzerland\\
$^5$M. R. \v{S}tef\'anik Observatory and Planetarium, Sl\'adkovi\v{c}ova 41, 920 01 Hlohovec, Slovak Republic\\
$^6$ Universit\'a degli studi di Napoli \qm{Federico II}, Dipartimento di Fisica \qm{Ettore Pancini}, Complesso Universitario di Monte S. Angelo, Via Cintia Edificio 6, 80126 Napoli, Italy\\
$^7$ Istituto Nazionale di Fisica Nucleare, Sezione di Napoli, Complesso Universitario di Monte S. Angelo, Via Cintia Edificio 6, 80126 Napoli, Italy\\
$^8$ INAF -- Osservatorio Astronomico di Roma,  Via Frascati, 33, Monteporzio Catone, 00078  Roma, Italy}

\date{\today}

\begin{abstract}
In this paper we investigate the three-dimensional (3D) motion of a test particle in a stationary, axially symmetric spacetime around a central compact object, under the influence of a radiation field. To this aim we extend the two-dimensional (2D) version of the Poynting-Robertson effect in General Relativity (GR) that was developed in previous studies. The radiation flux is modeled by photons which travel along null geodesics in the 3D space of a Kerr background and are purely radial  with respect to the zero angular momentum observer (ZAMO) frames. The 3D general relativistic equations of motion that we derive are consistent with the classical (i.e. non-GR) description of the Poynting-Robertson effect in 3D. The resulting dynamical system admits a critical hypersurface, on which radiation force balances gravity. Selected test particle orbits are calculated and displayed, and their properties described. It is found that test particles approaching the critical hypersurface at a finite latitude and with non-zero angular moment are subject to a latitudinal drift and asymptotically reach a circular orbit on the equator of the critical hypersurface, where they remain at rest with respect to the ZAMO. On the contrary, test particles that have lost all their angular momentum by the time they reach the critical hypersurface do not experience this latitudinal drift and stay at rest with respects to the ZAMO at fixed non-zero latitude. 
\end{abstract}

\maketitle
\section{Introduction}
\label{sec:intro}
Matter motion in a gravitational field may be affected, among others, by radiation forces. The case in which both the gravity and radiation fields originate in the same body has been discussed extensively in astrophysical context for decades. The Eddington argument, for instance, describes the outward radial force exerted by momentum transfer by radiation from a star and determines the conditions under which such force balances the inward gravitational force. The corresponding critical luminosity of a star separates the regime of radial infall from that of radial escape. It has long been known also that radiation can remove angular momentum from matter in motion around a star. Radiation absorbed by a particle is in general re-emitted isotropically in the reference frame of the particle. However re-emission in the star reference frame is (slightly) beamed along the direction of motion owing to relativistic aberration, such that the particle recoils opposite to its velocity and a fraction of its angular momentum is transferred to the radiation field. This effect was first studied by Poynting in 1903 \cite{Poynting1903}, based on a classical treatment, and extended to special relativity by Robertson in 1937 \cite{Robertson1937}; it has since then been termed Poynting-Robertson (PR) effect (or radiation drag). Both authors had in mind applications to the motion of comets, dusts, and cm-size bodies in the solar system, for which Newtonian gravity suffices \footnote{Note however that Robertson mentions general relativistic corrections to inertial terms in describing the perihelion shift in the quasi-Newtonian orbits \cite{Robertson1937}.}. The Sun is treated as a point-like source of gravity and radiation in a flat Minkowski spacetime, with straight light rays propagating outwards; the test particle moves in a planar orbit and reradiates energy at the same rate at which it receives it from the radiation field. Special relativistic equations are written in the test particle reference frame and then transformed to the reference system of the Sun \cite{Robertson1937}. More advanced descriptions of test particle motion under the influence of PR effects were discussed in a some classical works in the 50 -- 60's \citep{Wyatt1950,Guess1961}. A detailed description of the PR effect in relation to other radiation forces acting on small particles in the solar system was given in \cite{Burns1979}.

The range of applications of PR drag to astrophysical problems has grown steadily since 80's coming to encompass also compact objects, especially neutron stars (NSs) and black holes (BHs) that accrete matter down to the very strong gravitational fields in their vicinity. For instance Walker et al. \cite{Walker1989,Walker1992} studied the increase in mass accretion rate that is caused by PR drag when a bright thermonuclear flash occurs on the surface of a NS. This has motivated theory developments involving the PR effect in which GR is taken into account. The radial motion of test particles under the influence of a central isotropically emitting star was first investigated in the full GR in \cite{Abramowicz1990}, and detailed calculations in the Schwarzschild metric presented. General relativistic equations of motion for a fluid in an arbitrary radiation field were formulated in \cite{Carroll1990}. Miller and collaborators \cite{Miller1996,Miller1998} investigated the velocity field of accreting matter as affected by the PR effect in the vicinity of a rotating NS, by carrying out approximate calculations in the Kerr metric. 

A fully general relativistic treatment of the PR effect in the context of stationary and axially symmetric spacetimes was developed by Bini and collaborators \cite{Bini2009,Bini2011}. Similar to the classic model of Robertson these authors consider a compact object radiating as a point-like source, with photons traveling along null geodesics of the background spacetime (Schwarzschild or Kerr), and test particles moving in the equatorial plane around the compact object. Equations of motion are written in the ZAMO frame and then transformed to the rest reference frame of the compact object. The \emph{relativity of observer splitting formalism} is adopted, which permits to clearly distinguish between gravitational and inertial contributions (see \cite{Jantzen1992,Bini1997a,Bini1997b,Bini1998,Bini1999,Defalco2018}, for further details). The equations are solved numerically and test particle trajectories and motion analysed for both a purely radial photon field (null impact parameter) \cite{Bini2009} and a photon field endowed with angular momentum (non-null impact parameter) \cite{Bini2011}.   

Recent theoretical works on the extension of PR drag in GR have included studies of: test particle motion in the Vaidya spacetime \cite{Bini2011v} and around a slowly rotating relativistic star emitting isotropic radiation \cite{Oh2010}; the general relativistic PR effect on a spinning test particle \cite{Bini2010}; finite size effects \cite{Oh2011}, and the Lagrangian formulation of the general relativistic PR effect \cite{Defalco2018}. More astrophysically-oriented studies of PR effect in strong gravitational fields have concentrated on: the development of the {\it Eddington capture sphere concept} around luminous stars, the surface where gravity, radiation, and PR forces balance \cite{Wielgus2012,Stahl2012,Stahl2013,Wielgus2016,Wielgus2016n}; the {\it cosmic battery model} in astrophysical accretion discs \cite{Koutsantoniou2014,Contopoulos2015}; the dynamical evolution of accretion discs suddenly invested by a constant radiation filed (Bakala et al., 2018, A\&A submitted, \cite{Lancova2017}). Research in this area has acquired further momentum from the growing body of observational evidence for PR effect in matter motion around compact objects, especially accreting NSs undergoing thermonuclear flashes \cite{Ballantyne2004,Ballantyne2005,Worpel2013,Ji2014,Keek2014,Worpel2015,Keek2018}. 

Virtually all previous works on the general relativistic properties of the PR effect have been based upon a 2D model of the effect, i.e. planar (and arbitrarily oriented) orbits in spherically symmetric spacetimes (e.g. Schwarzschild's) and equatorial orbits in the (axially symmetric) Kerr metric. A necessary improvement consists in developing the 3D theory of the PR effect in GR. That would allow to investigate the motion of test particles immersed in non-spherically symmetric radiation fields (e.g. latitude-dependent fields) and/or orbiting away from the equatorial plane of the Kerr metric. That is the aim of the present study, which builds on the formalism developed in Refs. \cite{Bini2009,Bini2011}. Our paper is structured as follows: in Sec. \ref{sec:GRPR} we generalise to the 3D case the previous 2D equations for the PR effect in a stationary and axially symmetric general relativistic spacetimes. We adopt a simple prescription for the radiation, namely a field with zero angular momentum. In Sec. \ref{sec:critc_rad} we define the critical hypersurface on which radiation force balances gravity and discuss its salient features.  In Sec. \ref{sec:orbits} we present calculations of selected orbits in the Schwarzschild and Kerr spacetimes; our concluding remarks are in Sec. \ref{sec:end}. 

\section{Scenario and spacetime geometry}
\label{sec:GRPR}
\begin{figure*}[t]
\centering
\includegraphics[scale=0.6]{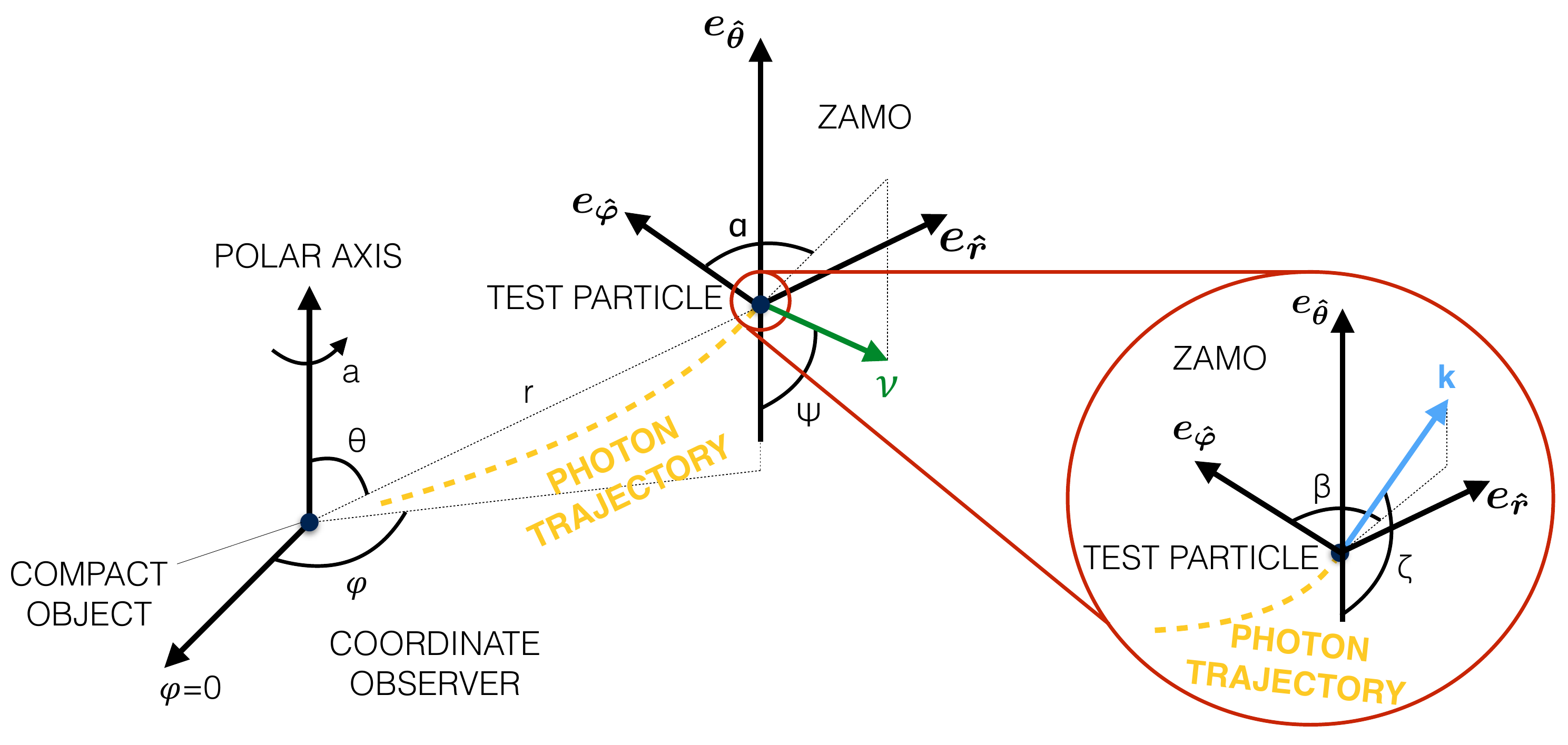}
\caption{Visual representation of the radiation field-test particle interaction geometry in the Kerr metric. The spatial location of the test particle is described by Boyer-Linquist coordinates $\left\{\boldsymbol{r},\boldsymbol{\theta},\boldsymbol{\varphi}\right\}$. The ZAMO local frame is $\left\{\boldsymbol{e_{\hat{t}}},\boldsymbol{e_{\hat{r}}},\boldsymbol{e_{\hat{\theta}}},\boldsymbol{e_{\hat{\varphi}}}\right\}$. The photons of the radiation field travel along null geodesics of the background spacetime with four-momentum $\boldsymbol{k}$.  Two photon impact parameters, $b$ and $q$ are related respectively to the two angles $\beta$ and $\zeta$, formed in the local ZAMO frame. The test particle moves in the 3D space with a velocity $\nu$, forming the azimuthal, $\alpha$, and polar, $\psi$, angles in the local ZAMO frame.}
\label{fig:Fig1}
\end{figure*}

Our scenario for the description of the interaction between the radiation field and the motion of a test particle in the extreme gravitational field of a BH, or a NS, is constituted as follows: we consider the radiation field coming from an emitting region, located outside of the event horizon. The test particle motion is determined by its position in spherical coordinates and its velocity field in the ZAMO frame. The photon four-momentum is described by a pair of polar coordinates (see Fig. \ref{fig:Fig1}). In order to derive such set of equations we compute first the quantities in the ZAMO frame and then we transform them in the static observer frame. To deal with the relative motion of two non-inertial observers in GR we use the relativity of observer splitting formalism.

We consider a central compact object (BH or NS), whose outside spacetime is described by the Kerr metric with signature $(-,+,+,+)$ \cite{Kerr1963}. In geometrical units ($c = G = 1$), the line element of the Kerr spacetime, $ds^2=g_{\alpha\beta}dx^\alpha dx^\beta$, in Boyer-Lindquist coordinates, parameterized by mass $M$ and spin $a$, reads as \cite{Boyer1967}
\begin{equation}\label{kerr_metric}
\begin{aligned}
\mathrm{d}s^2 &=-\left(1-\frac{2Mr}{\Sigma}\right) \,\mathrm{d}t^2-\frac{4Mra}{\Sigma} \sin^2\theta\,\mathrm{d}t\, \mathrm{d}\varphi\\
&+\frac{\Sigma}{\Delta}\, \mathrm{d}r^2+\Sigma \,\mathrm{d}\theta^2+\rho\sin^2\theta\, \mathrm{d}\varphi^2, 
\end{aligned}
\end{equation}
where $\Sigma \equiv r^{2} + a^{2}\cos^{2}\theta$, $\Delta \equiv r^{2} - 2Mr + a^{2}$, and $\rho  \equiv r^2+a^2+2Ma^2r\sin^2\theta/\Sigma$. The determinant of the Kerr metric is $g=\sqrt{(\Delta/\rho)g_{rr}g_{\varphi\varphi}g_{\theta\theta}}\equiv-\Sigma^2\sin^{2}\theta$.

\subsection{ZAMO frame}
ZAMOs are dragged by the rotation of the spacetime with angular velocity $\Omega_{\mathrm{ZAMO}}=-g_{\phi t}/g_{\phi \phi}$, while their radial coordinate remains constant. The four-velocity $\boldsymbol{n}$ of ZAMOs is the future-pointing unit normal to the spatial hypersurfaces, i.e. \cite{Bini1997a,Bini1997b,Bini2009,Bini2011},
\begin{equation}
\label{n}
\boldsymbol{n}=\frac{1}{N}(\boldsymbol{\partial_t}-N^{\varphi}\boldsymbol{\partial_\varphi})\,,
\end{equation}
where $N=(-g^{tt})^{-1/2}$ is the time lapse function, $g^{tt}=g_{\varphi\varphi}/(g_{tt}g_{\varphi\varphi}-g_{t\varphi}^2)$, and $N^{\varphi}=g_{t\varphi}/g_{\varphi\varphi}$ the spatial shift vector field. We focus our attention on the region outside the event horizon, where the time coordinate hypersurfaces are spacelike, i.e., $g^{tt}<0$. An orthonormal frame adapted to the ZAMOs is given by \cite{Bardeen1972}
\begin{equation} \label{eq:zamoframes}
\begin{aligned}
&\boldsymbol{e_{\hat t}}=\boldsymbol{n},\quad
\boldsymbol{e_{\hat r}}=\frac1{\sqrt{g_{rr}}}\boldsymbol{\partial_r},\\
&\boldsymbol{e_{\hat \theta}}=\frac1{\sqrt{g_{\theta \theta }}}\boldsymbol{\partial_\theta},\quad
\boldsymbol{e_{\hat \varphi}}=\frac1{\sqrt{g_{\varphi \varphi }}}\boldsymbol{\partial_\varphi}.
\end{aligned}
\end{equation}
The relative dual tetrad of one-forms is given by
\begin{equation} \label{eq:zamoframed}
\begin{aligned}
&\boldsymbol{\omega^{{\hat t}}}= N dt,\quad \boldsymbol{\omega^{{\hat r}}} =\sqrt{g_{rr}} dr,\\
&\boldsymbol{\omega^{{\hat \theta}}}= \sqrt{g_{\theta \theta }} d\theta,\quad
\boldsymbol{\omega^{{\hat \varphi}}}=\sqrt{g_{\varphi \varphi }}(d \varphi+N^{\varphi} dt).
\end{aligned}
\end{equation}
All the indices associated to the ZAMO frame will be labeled by a hat, instead all the quantities measured in the ZAMO frame will be followed by $(n)$. 

\subsection{ZAMO kinematical quantities\\ }
Since the accelerated ZAMOs are locally nonrotating, their vorticity vector $\boldsymbol{\omega}(n)$ vanishes, but they have a nonzero expansion tensor $\boldsymbol{\theta}(n)$. For this reason it is more convenient to use the Lie transport (see \cite{Bini1997a,Defalco2018}, for further details). The nonzero ZAMO kinematical quantities (i.e., acceleration $\boldsymbol{a}(n)=\nabla_{\boldsymbol{n}} \boldsymbol{n}$, expansion tensor along the $\hat{\varphi}$-direction $\boldsymbol{\theta_{\hat\varphi}}(n)$, also termed shear vector, and the relative Lie curvature vector $\boldsymbol{k}_{(\rm Lie)}(n)$) have only nonzero components in the $\hat{r}-\hat{\theta}$ plane of the tangent space \cite{Bini2009,Bini2011,Defalco2018}:
\begin{equation}\label{accexp}
\begin{aligned}
\boldsymbol{a}(n) & = a(n)^{\hat r}\, \boldsymbol{e_{\hat r}} + a(n)^{\hat\theta}\, \boldsymbol{e_{\hat\theta}}\\
&=\partial_{\hat r}(\ln N)\, \boldsymbol{e_{\hat r}} + \partial_{\hat\theta}(\ln N)\, \boldsymbol{e_{\hat\theta}},\\
\boldsymbol{\theta_{\hat\varphi}}(n)& = \theta(n)^{\hat r}{}_{\hat\varphi}\, \boldsymbol{e_{\hat r}} + \theta(n)^{\hat\theta}{} _{\hat\varphi}\, \boldsymbol{e_{\hat \theta}}\\
&= -\frac{\sqrt{g_{\varphi\varphi}}}{2N}\,(\partial_{\hat r} N^\varphi\, \boldsymbol{e_{\hat r}} + \partial_{\hat\theta} N^\varphi\, \boldsymbol{e_{\hat \theta}}),\\
\boldsymbol{k}_{(\rm Lie)}(n)& =  k_{(\rm Lie)}(n)^{\hat r}\, \boldsymbol{e_{\hat r}} + k_{(\rm Lie)}(n)^{\hat\theta}\, \boldsymbol{e_{\hat\theta}}\\
& = -[\partial_{\hat r}(\ln \sqrt{g_{\varphi\varphi}})\, \boldsymbol{e_{\hat r}} + \partial_{\hat\theta}(\ln \sqrt{g_{\varphi\varphi}})\, \boldsymbol{e_{\hat\theta}}].
\end{aligned}
\end{equation}
In Table \ref{tab:ZAMOq} we summarize the expressions of such quantities for the Kerr spacetime. 
\renewcommand{\arraystretch}{1.8}
\begin{table*}
\begin{center}
\caption{\label{tab:ZAMOq} Explicit expressions of metric and ZAMO kinematical quantities for the Kerr metric.}	
\normalsize
\begin{tabular}{l  c} 
\ChangeRT{1pt}
{\bf Metric quantity} & {\bf Explicit expression} \\
\ChangeRT{1pt}
$N=(-g^{tt})^{-1/2}$ & $[\Delta /\rho]^{1/2}$   \\
\hline
$N^\varphi=g_{t\varphi}/g_{\varphi\varphi}$ & $-2Mar/[\Sigma \rho]$ \\
\ChangeRT{1pt}
{\bf ZAMO quantity} & {\bf Explicit expression} \\
\ChangeRT{1pt}
\hline
\hline
\multicolumn{2}{c}{\emph{Radial components}}\\
\hline
\hline
$a(n)^{\hat r}$& $M/[\rho\sqrt{\Sigma^5\Delta}]\left\{\Sigma^2(r^2-a^2)+a^2\sin^2\theta[r^2(3r^2-4Mr+a^2)+a^2\cos^2\theta(r^2-a^2)]\right\}$   \\
\hline
$\theta(n)^{\hat r}{}_{\hat\varphi}$& $aM\sin\theta[(r^2+a^2)(\Sigma-2r^2)-2r^2\Sigma]/[\rho\sqrt{\Sigma^5}]$\\
\hline
$k_{(\rm Lie)}(n)^{\hat r}$ & $-\sqrt{\Delta/\Sigma^5}[r\Sigma^2+a^2M\sin^2\theta(\Sigma-2r^2)]/\rho$\\
\hline
\hline
\multicolumn{2}{c}{\emph{Polar components}}\\
\hline
\hline
$a(n)^{\hat\theta}$ & $-a^2rM\sin(2\theta)[r^2+a^2]/[\rho\sqrt{\Sigma^5}]$\\
\hline
$\theta(n)^{\hat\theta}{} _{\hat\varphi}$ & $a^2rM\sin(2\theta)\sin\theta\sqrt{\Delta}/[\rho\sqrt{\Sigma^5}]$\\
\hline
$k_{(\rm Lie)}(n)^{\hat\theta}$ & $-\sin(2\theta)[(r^2+a^2)(2a^2rM\sin^2\theta+\Sigma^2)+2a^2rM\Sigma\sin^2\theta]/[2\rho\sqrt{\Sigma^5}\sin^2\theta]$\\
\ChangeRT{1pt}
\end{tabular}
\end{center}
\end{table*}

\subsection{Radiation field}
\label{sec:phot}
The stress-energy tensor, describing the radiation field, is modeled as a coherent flux of photons traveling along null geodesics in the Kerr geometry and acting on the test particle in the following manner \cite{Robertson1937,Bini2009,Bini2011}
\begin{equation}
\label{STE}
T^{\mu\nu}=\Phi^2 k^\mu k^\nu\,,\qquad k^\mu k_\mu=0,\qquad k^\mu \nabla_\mu k^\nu=0,
\end{equation}
where parameter $\Phi$ is related to the intensity of the radiation field and $\boldsymbol{k}$ is the four-momentum field describing the null geodesics. The photon four-momentum, $\boldsymbol{k}$, and the photon spatial unit relative velocity with respect to the ZAMOs, $\boldsymbol{\hat{\nu}}(k,n)$, are respectively given by
\begin{equation} \label{photon}
\begin{aligned}
&\boldsymbol{k}=E(n)[\boldsymbol{n}+\boldsymbol{\hat{\nu}}(k,n)],\\
&\boldsymbol{\hat{\nu}}(k,n)=\sin\zeta\sin\beta\ \boldsymbol{e_{\hat r}}+\cos\zeta\ \boldsymbol{e_{\hat\theta}}+\sin\zeta \cos\beta\ \boldsymbol{e_{\hat\varphi}},
\end{aligned}
\end{equation}
where $\beta$ and $\zeta$ are the two angles in the azimuthal and polar direction, respectively (see Fig. \ref{fig:Fig1}). The case of $\sin \beta >0$ corresponds to an outgoing photon beam (increasing $r$) while the case of $\sin \beta <0$ corresponds to an incoming photon beam (decreasing $r$, see Fig. \ref{fig:Fig1}). The photon four-momentum in the background Kerr geometry is identified by two impact parameters $(b,q)$, which are associated with two emission angles $(\beta,\zeta)$, respectively.

Using Eq. (\ref{photon}), the photon energy with respect to the ZAMO, $E(n)$, is expressed in the frame of a distant static observer by
\begin{equation} \label{energyZAMO}
\begin{aligned}
E(n)&=-\boldsymbol{k}(n)\cdot \boldsymbol{n}=-\boldsymbol{k}\cdot\frac{1}{N}\left(\boldsymbol{\partial_t}-N^{\varphi}\boldsymbol{\partial_\varphi}\right)\\
&=\frac{E+L_zN^\varphi}{N}=\frac{E}{N}(1+bN^\varphi),
\end{aligned}
\end{equation}
where $E=-k_t>0$ is the conserved photon energy, $L_z=k_\varphi$ is the conserved angular momentum along the polar $z$ axis orthogonal to the equatorial plane, and $b\equiv -k_{\phi}k_{t}=L_z/E$ is the first (azimuthal) photon impact parameter (constant of motion) \cite{Carter1968}; note that all these quantities are measured by a distant static observer \cite{Bini2011}. 

This impact parameter is associated with the relative azimuthal angle $\beta$, measured in the ZAMO frame \cite{Bini2011} (see Fig. \ref{fig:Fig1}). The angular momentum along the polar $\hat{\theta}$-axis in the ZAMO frame, $L_z(n)$, is expressed in the distant static observer frame by
\begin{equation} \label{ang1}
\begin{aligned}
E(n)\cos\beta\sin\zeta&=L_z(n)=\boldsymbol{k}(n)\cdot\boldsymbol{e_{\hat \varphi}}\\
&=\boldsymbol{k}\cdot\frac{\boldsymbol{\partial_\varphi}}{\sqrt{g_{\varphi\varphi}}}=\frac{L_z}{\sqrt{g_{\varphi\varphi}}}
\end{aligned}
\end{equation}
From such equation, we obtain
\begin{equation} \label{ANG1}
\begin{aligned}
\cos\beta&=\frac{b E}{\sin\zeta\sqrt{g_{\varphi\varphi}}E(n)}=\frac{L_zN}{\sin\zeta\sqrt{g_{\varphi\varphi}}(E+L_zN^\varphi)}\\
&=\frac{b N}{\sin\zeta\sqrt{g_{\varphi\varphi}}(1+b N^\varphi)}.
\end{aligned}
\end{equation}
An equation for $\zeta$ is needed to completely determine $\beta$.

The photon specific four-momentum components in the Kerr geometry are given by \cite{Chandrasekhar1992}
\begin{align}
     k^{t} &= \Sigma^{-1} \left(a\,b-a^2\sin^{2}\theta+(r^2+a^2)P\,\Delta^{-1}\right)\,,\nonumber\\
     k^{r} &=  s_r\Sigma^{-1} \sqrt{R_{b,q}(r)}\,,\label{CarterEQs}\\
     k^{\theta} &=  s_{\theta}\Sigma^{-1} \sqrt{\Theta_{b,q}(\theta)}\,,\nonumber\\
     k^{\varphi} &=  \Sigma^{-1}\left(b\,\mathrm{cosec}^{2} \theta-a+a\,P\,\Delta^{-1}\right)\,,\nonumber
\end{align} 
where $P\equiv r^{2} + a^{2}-b\,a$, and the pair of signs $s_{r}$, $s_{\theta}$ describes the orientation of the radial and latitudinal evolution, respectively \cite{Carter1968}. The radial and latitudinal effective potentials are respectively \cite{Chandrasekhar1992}:
\begin{align}
       R_{b,q} \left( r \right) &=  \left( r^{2} + a^{2}
 - a b \right) ^{2}
       - \Delta \left[ q + \left( b - a \right) ^{2} \right]\,,\label{Rpot} \\
       \Theta_{b,q} \left( \theta \right) &= q + a^{2} \cos^{2}
\theta -b^{2} \mathrm{cot}^{2} \theta\,.\label{Thetapot}
\end{align} 
Here, $q$ is the second (latitudinal) photon impact parameter (constant of motion) related to the covariant components of the photon four-momentum through the relation \cite{Chandrasekhar1992}
 \begin{equation} \label{q_def2}
q \equiv \left(\frac{k_{\theta}}{k_{t}}\right)^{2} + \left[b\tan\left(\frac{\pi}{2} - \theta\right)\right]^{2} - a^{2}\cos^{2}\theta\,.    
 \end{equation}

\subsubsection{Impact parameters}
We consider here a radiation field which consists of photons moving in a purely radial direction at the infinity (this physically admissible because of the asymptotic flatness of the Kerr spacetime). In this case we have
\begin{equation}
\boldsymbol{k}=\boldsymbol{\partial_t}+\boldsymbol{\partial_r}\,. \label{infty_mom}
\end{equation}
From Eq. (\ref{infty_mom}) we note that the azimuthal component of the four-momentum, $k^\varphi$, vanishes at infinity. Moreover, in order to simplify the calculations we assume that the azimuthal impact parameter of the radiation field, $b$, takes null value, i.e., $b = 0$. The latitudinal impact parameter, $q$, can be calculated from the condition
\begin{equation} \label{cond2}
\Theta_{b=0,q}( \theta)=0,
\end{equation}
which results from the absence of latitudinal photon motion ($k^{\theta}=0$). From Eqs. (\ref{Thetapot}) and (\ref{cond2}) we can express  $q$ as a function of the polar angle $\theta$:
\begin{equation} \label{q_r}
q=- a^{2} \cos^{2}\theta\,. 
\end{equation}
This is possible because the latitudinal potential, Eq. (\ref{Thetapot}), is independent of the radial coordinate and therefore the polar angle $\theta$ along a given photon trajectory is conserved.  Photons with a given value of $q$ move only in the radial and azimuthal directions on the surface of the cone with the vertex located in the coordinates origin and with the vertex angle  $\theta$ given by Eq. (\ref{q_r}). Note that the azimuthal motion on finite values of the radial coordinate is caused only by frame dragging.

The above-defined radiation field significantly simplifies the integration of test particle trajectories in that only a single photon beam, described by the constants of motion $b=0,\ q=- a^{2} \cos^{2}\theta$, must be considered at the test particle position. In such case, the radial potential, Eq. ({\ref{Rpot}}), is always positive above the event horizon: this proves that the radiation field reaches every positions (for all $r$ and $\theta$) above the event horizon. The second constant of motion $q$ ranges in the interval $[-a^2, 0]$.  The value $q=0$ corresponds to the motion of photons in the equatorial plane, while the value of $q=  -a^2$ corresponds to the motion of photons along the polar axis on the south or north directions. We note also that since $q \leq 0$ radiation field photons can never cross the equatorial plane.

The  local components of the photon four-momentum in the ZAMO frame are obtained through the following transformation:
\begin{equation}
k^{\hat{\mu}} = \omega^{\hat{\mu}}_{\,\,\,\alpha}\,k^{\alpha},
\end{equation}
where $\omega^{\hat{\mu}}_{\,\,\,\alpha}$ represents the transformation matrix from the holonomic basis $\boldsymbol{\partial_\alpha}$ to the anholomic (tetaraed) basis $\boldsymbol{e_{\hat\alpha}}$, see Eq. (\ref{eq:zamoframes}) for determining its components. The local polar direction $\zeta$ of the photon four-momentum is given by (see Fig. \ref{fig:Fig1})
\begin{equation} \label{angle1}
\cos \zeta =-\frac{k^{\hat{\theta}} }{k^{\hat{t} }}.
\end{equation}
For the considered radiation field ($q=- a^{2} \cos^{2}\theta\,,k^{\theta}=0$) we simply obtain
\begin{equation} \label{angle2}
k^{\hat{\theta}} = \omega^{\hat{\theta}}_{\,\,\,\theta}\,k^{\theta}=0.
\end{equation}
Consequently from Eqs. (\ref{angle1}) and (\ref{angle2}), the local polar direction of the radiation field photons in the ZAMO frame is always $\zeta=\pi/2$. From Eq. (\ref{ang1}) and $b=0$, the local azimuthal direction $\beta$ of the photon four-momentum is $\cos\beta=0$ (see Fig. \ref{fig:Fig1}). Therefore, the local azimuthal angle of the test field photons in the  ZAMO frame always take the value of $\beta=\pi/2$. We can conclude that in all ZAMO frames radiation field photons move in a purely radial direction. 
The source of the radiation field can thus be considered as centered in the coordinate origin, (differentially) rotating with a latitude-dependent angular velocity $\Omega_{\mathrm{ZAMO}}$ and emitting photons only along the radial direction in the appropriate, locally comoving ZAMO frame. 

\subsubsection{Intensity  parameter}
Since the photon four-momentum $\boldsymbol{k}$ is completely determined by  $(b\,,q)$, the coordinate dependence of $\Phi$ then follows from the conservation equations $\nabla_{\beta}T^{\alpha\beta}=0$. Exploiting the absence of photon latitudinal motion ($k^{\theta}=0$) and symmetries of the Kerr spacetime, these can be written as
\begin{equation}
\label{flux_cons}
\begin{aligned}
0&=\nabla_\beta (\Phi^2 k^\beta)=\frac{1}{\sqrt{-g}}\partial_\beta (\sqrt{-g}\,\Phi^2 k^\beta)\\
&=\partial_r(\sqrt{-g}\,\Phi^2 k^r).
\end{aligned}
\end{equation}
Therefore, we have 
\begin{equation}
\label{cons_condition}
\begin{aligned}
\sqrt{-g}\,\Phi^2 k^r&=NE(n)\sqrt{g_{\varphi\varphi}g_{\theta\theta}}\sin\zeta\sin\beta\\
&=\hbox{const}=E\Phi^2_0,  
\end{aligned}
\end{equation}
where $\Phi_0$ is a new constant related to the intensity of the radiation field at the emitting surface. This equation, however, does not fix the intensity parameter unambiguously. In fact, the conservation equations will be fulfilled, even if we multiply the constant expression $E\Phi_0^2$ by an arbitrary function of the $\theta$ coordinate. Thus this condition determines the class of radiating fields that differ from one another by the latitudinal dependence of the intensity. A radiation field which is independent of latitude (and whole intensity parameter is thus independent of $\theta$) is a natural choice, especially in the Schwarzschild limit because of its spherical symmetry. This can easily achieved by multiplying (\ref{cons_condition}) by a factor $\sin\theta$, such that the intensity parameter becomes 
\begin{equation} \label{eq:int}
\Phi^2=\frac{\Phi_0^2\sin\theta}{\sqrt{g_{\varphi\varphi}g_{\theta\theta}}}\equiv \frac{\Phi_0^2}{\sqrt{(r^2+a^2)^2-a^2\,\Delta\,\sin^{2}\theta}},
\end{equation}
where we have used Eqs. (\ref{ang1}) and (\ref{ANG1}), together with the fact that  $N|b\tan\beta|=\sin\zeta\sqrt{g_{\varphi\varphi}}$ for $b=0$. In a Schwarzschild spacetime limit Eq. (\ref{eq:int}) thus reads
\begin{equation}
\Phi^2=\frac{\Phi_0^2}{r^2}\,,
\end{equation}
which matches the spacetime spherical symmetry. 

\subsection{Test particle motion}
\label{sec:test_part}
We consider a test particle moving in the 3D space, with four-velocity $\bold{U}$ and spatial three-velocity with respect to the ZAMOs, $\boldsymbol{\nu}(U,n)$:
\begin{eqnarray} 
\boldsymbol{U}&=&\gamma(U,n)[\boldsymbol{n}+\boldsymbol{\nu}(U,n)], \label{testp1}\\
\boldsymbol{\nu}(U,n)&=&\nu^{\hat r}\ \boldsymbol{e_{\hat r}}+\nu^{\hat\varphi}\ \boldsymbol{e_{\hat \varphi}}+\nu^{\hat\theta}\ \boldsymbol{e_{\hat\theta}} \label{testp2}\\ 
&=&\nu\sin\psi\sin\alpha\ \boldsymbol{e_{\hat r}}+\nu\cos\psi\ \boldsymbol{e_{\hat\theta}}+\nu\sin\psi \cos\alpha\ \boldsymbol{e_{\hat\varphi}}\nonumber,\\ \nonumber
\end{eqnarray}
where $\gamma(U,n)=1/\sqrt{1-||\boldsymbol{\nu}(U,n)||^2}$ is the Lorentz factor (see Fig. \ref{fig:Fig1}). We use the following abbreviated notations $\nu^{\hat \alpha}=\nu(U,n)^{\hat \alpha}$, $\nu=||\boldsymbol{\nu}(U,n)||\ge0$, $\gamma(U,n) =\gamma$ throughout this paper. We have that $\nu$ represents the magnitude of the test particle spatial velocity $\boldsymbol{\nu}(U,n)$, $\alpha$ is the azimuthal angle of the vector $\boldsymbol{\nu}(U,n)$ measured clockwise from the positive $\hat \varphi$ direction in the $\hat{r}-\hat{\varphi}$ tangent plane in the ZAMO frame, and $\psi$ is the polar angle of the vector $\boldsymbol{\nu}(U,n)$ measured from the axis orthogonal to the $\hat{r}-\hat{\varphi}$ tangent plane of the ZAMO frame (see Fig. \ref{fig:Fig1}). The explicit expression for the test particle velocity components with respect to the ZAMOs are \cite{Bini2009,Bini2011}:
\begin{equation} 
\begin{aligned}
&U^t\equiv \frac{dt}{d\tau}=\frac{\gamma}{N},\quad U^r\equiv \frac{dr}{d\tau}=\frac{\gamma\nu^{\hat r}}{\sqrt{g_{rr}}},\\
&U^\theta\equiv \frac{d\theta}{d\tau}=\frac{\gamma\nu^{\hat\theta}}{\sqrt{g_{\theta\theta}}},\quad U^\varphi\equiv \frac{d\varphi}{d\tau}=\frac{\gamma\nu^{\hat\varphi}}{\sqrt{g_{\varphi\varphi}}}-\frac{\gamma N^\varphi}{N},
\end{aligned}
\end{equation}
where $\tau$ is the proper time parameter along $\bold{U}$. 

\subsubsection{Relativity of observer splitting formalism}
\label{sec:rosf} 
The acceleration of the test particle relative to the ZAMO congruence, $\boldsymbol{a}(U)=\nabla_{\bold U} \bold{U}$, is given  by the formula (see Eq. (29) in \cite{Defalco2018} and references therein) \footnote{A complementary approach to the relativity of observer splitting formalism is the general relativistic Lagrangian formulation of the PR effect \cite{Defalco2018}.}:
\begin{equation} \label{tpacc}
\begin{aligned}
a(U)^\alpha&=\gamma^2\left[a(n)^\alpha+\Gamma(n)^\alpha{}_{\beta\gamma}\nu(U,n)^\beta\nu(U,n)^\gamma\right.\\
&\left.+2\theta(n)^\alpha{}_\beta\nu(U,n)^\beta\right]+\frac{d(\gamma\nu(U,n)^\alpha)}{d\tau},
\end{aligned}
\end{equation}
where $\alpha,\beta,\gamma=\hat{r},\hat{\theta},\hat{\varphi}$ run on the spatial indices of the metric coordinates \footnote{
Terms $C_{\rm(Lie)}(n)^\alpha_{\beta\gamma},\ C_{\rm(Lie)}(n)^\alpha_\beta$, representing respectively the temporal and spatial constant structures, are missing in Eq. (\ref{tpacc}), because they vanish in a stationary and axially-symmetric spacetime (see \cite{Bini1997a,Bini1997b,Defalco2018}, for details).}. Calculating the Christoffel symbols $\Gamma(n)^\alpha{}_{\beta\gamma}$, we have \cite{Bini1997a,Bini1997b,Defalco2018}
\begin{equation}
\begin{aligned}
\Gamma(n)^{\hat r}{}_{\hat \varphi \hat\varphi}&=\Gamma(n)^{\hat r}{}_{\hat \theta \hat\theta}=-2\Gamma(n)^{\hat \varphi}{}_{\hat r\hat\varphi}\\
&=-2\Gamma(n)^{\hat \theta}{}_{\hat r \hat\theta}=k_{\rm (Lie)}(n)^{\hat r},\\
-2\Gamma(n)^{\hat \varphi}{}_{\hat \varphi\hat\theta}&=\Gamma(n)^{\hat \theta}{}_{\hat \varphi\hat\varphi}=k_{\rm (Lie)}(n)^{\hat \theta}. 
\end{aligned}
\end{equation}
Therefore, Eqs. (\ref{tpacc}) in explicit form become
\begin{eqnarray}
a(U)^{\hat r}&=& \gamma^2 [a(n)^{\hat r}+k_{\rm (Lie)}(n)^{\hat r}\,\nu^2 (\cos^2\alpha\sin^2\psi\label{acc1} \\
&&+\cos^2\psi)+2\nu\cos \alpha\sin\psi\, \theta(n)^{\hat r}{}_{\hat \varphi}]\nonumber\\
&&+\gamma \left(\gamma^2 \sin\alpha\sin\psi \frac{\rm d \nu}{\rm d \tau}+\nu \cos \alpha\sin\psi \frac{\rm d \alpha}{\rm d \tau}\right.\nonumber\\
&&\left.+\nu \cos \psi\sin\alpha \frac{\rm d \psi}{\rm d \tau} \right), \nonumber\\   
a(U)^{\hat \theta}&=&\gamma^2 [a(n)^{\hat \theta}+k_{\rm (Lie)}(n)^{\hat \theta}\,\nu^2 \sin^2\psi\cos^2\alpha\label{acc3}\\
&&-k_{\rm (Lie)}(n)^{\hat r}\, \nu^2\sin\psi\sin\alpha\cos\psi\nonumber\\
&&+2\nu\cos \alpha\sin\psi\, \theta(n)^{\hat \theta}{}_{\hat \varphi}]\nonumber\\
&&+ \gamma\left(\gamma^2 \cos\psi \frac{\rm d \nu}{\rm d \tau}-\nu \sin\psi \frac{\rm d \psi}{\rm d \tau}\right).\nonumber\\
a(U)^{\hat \varphi}&=& -\gamma^2 \nu^2\cos \alpha\sin\psi\left[ \sin \alpha \sin\psi\, k_{\rm (Lie)}(n)^{\hat r}\right.\label{acc2}\\ 
&&\left.+ k_{\rm (Lie)}(n)^{\hat \theta}\cos\psi\right]+ \gamma\left(\gamma^2 \cos \alpha\sin\psi \frac{\rm d \nu}{\rm d \tau}\right.\nonumber\\
&&\left.-\nu\sin \alpha\sin\psi \frac{\rm d \alpha}{\rm d \tau}+\nu\cos\alpha\cos\psi \frac{\rm d \psi}{\rm d \tau}\right),\nonumber
\end{eqnarray}
From the orthogonality between $(\bold{a}(U),\bold{U})$ we have:
\begin{equation}\label{acc4}
\begin{aligned}
a(U)^{\hat t}&=\nu[a(U)^{\hat r}\sin\alpha\sin\psi+a(U)^{\hat \theta}\cos\psi\\
&+a(U)^{\hat \varphi}\cos\alpha\sin\psi]\\   
&=\gamma^2\nu\left\{\sin \alpha \sin\psi\left[a(n)^{\hat r}\right.\right.\\
&\left.\left.+2\nu\cos \alpha\sin\psi\, \theta(n)^{\hat r}{}_{\hat \varphi}\right] \right. \\
&\left.+\cos\psi\left[a(n)^{\hat \theta}\right.\right.\\
&\left.\left.+2\nu\cos\alpha\sin\psi \theta(n)^{\hat \theta}{}_{\hat \varphi}\right]\right\}+ \gamma^3\nu \frac{\rm d \nu}{\rm d \tau}. 
\end{aligned}         
\end{equation}

\subsection{Radiation test particle interaction}
\label{sec:eoms}
We assume that the interaction between the test particle and the radiation files takes place through Thomson scattering, characterized by a constant $\sigma$, independent of direction and frequency of the radiation field. The radiation force is \cite{Abramowicz1990,Bini2009,Bini2011}
\begin{equation} \label{radforce}
{\mathcal F}_{\rm (rad)}(U)^\alpha = -\sigma P(U)^\alpha{}_\beta \, T^{\beta}{}_\mu \, U^\mu \,,
\end{equation}
where $P(U)^\alpha{}_\beta=\delta^\alpha_\beta+U^\alpha U_\beta$ projects a vector orthogonally to $\bold{U}$, namely on the spatial hypersurfaces or local rest spaces. The test particle equations of motion then become $m \bold{a}(U) = \boldsymbol{{\mathcal F}_{\rm (rad)}}(U)$, where $m$ is the test particle mass. By definition the radiation force lies in the local rest space of the test particle; to calculated it we decompose the photon four-momentum $\bold{k}$ first with respect to the four-velocity of the test particle, $\bold{U}$, and then to the previous ZAMO decomposition, $\bold{n}$, i.e. \cite{Bini2009,Bini2011},
\begin{equation} \label{diff_obg}
\bold{k} = E(n)[\bold{n}+\boldsymbol{\hat{\nu}}(k,n)]=E(U)[\bold{U}+\boldsymbol{\hat {\mathcal V}}(k,U)].
\end{equation}
By projecting $\bold{k}$ with respect to the test particle four-velocity, $\bold{U}$ we get
\begin{equation} \label{helpsplit}
\bold{P}(U)\cdot\bold{k}=E(U)\boldsymbol{\hat {\mathcal V}}(k,U)\,,\quad 
\bold{U}\cdot \bold{k}=-E(U).
\end{equation}
Using Eq. (\ref{helpsplit}) in Eq. (\ref{radforce}) we obtain
\begin{equation} \label{Frad0}
\begin{aligned}
{\mathcal F}_{\rm (rad)}(U)^\alpha&=-\sigma \Phi^2 [P(U)^\alpha{}_\beta k^\beta]\, (k_\mu U^\mu)\\
&=\sigma \, [\Phi E(U)]^2\, \hat {\mathcal V}
(k,U)^\alpha.
\end{aligned}
\end{equation}
In this way the test particle acceleration is aligned with the photon relative velocity in the test particle local rest space, i.e.,
\begin{equation}\label{geom}
\bold{a}(U)=\tilde \sigma \Phi^2 E(U)^2  \,\boldsymbol{\hat {\mathcal V}}(k,U)\,,
\end{equation}
where $\tilde \sigma=\sigma/m$. Hereafter we use the simplified notation $\boldsymbol{\hat {\mathcal V}}(k,U)=\boldsymbol{\hat {\mathcal V}}$. Multiplying scalarly Eq. (\ref{diff_obg}) by $\bold{U}$ and using Eqs. (\ref{photon}) (i.e. the decomposition of $\bold{k}$ in the ZAMO frame), and (\ref{testp1}) -- (\ref{testp2}) (i.e. the decomposition of $\bold{U}$ in the ZAMO frame), we find
\begin{equation} \label{enepart}
\begin{aligned}
E(U)&=\gamma E(n)[1-\boldsymbol{\nu}(U,n)\boldsymbol{\hat\nu}(k,n)]\\
&=\gamma E(n)[1-\nu(\sin\zeta\sin\psi\cos(\alpha-\beta)\\
&+\cos\zeta\cos\psi)]\\
&=\gamma \frac{E}{N}[1-\nu\sin\psi\sin\alpha],
\end{aligned}
\end{equation}
where we have used Eqs. (\ref{ang1}) and (\ref{ANG1}) and the value of the assumed local angles. Such procedure is very useful for determining the spatial velocity $\boldsymbol{\hat{\mathcal{V}}}$:
\begin{equation}
\boldsymbol{\hat{\mathcal{V}}}=\left[\frac{E(n)}{E(U)}-\gamma\right]\boldsymbol{n}+\frac{E(n)}{E(U)}\boldsymbol{\hat{\nu}}(k,n)-\gamma\boldsymbol{\nu}(U,n).
\end{equation}
The frame components of $\boldsymbol{\hat{\mathcal{V}}}=\hat{\mathcal{V}}^t\boldsymbol{n}+\hat{\mathcal{V}}^r\boldsymbol{e_{\hat r}}+\hat{\mathcal{V}}^\theta \boldsymbol{e_{\hat\theta}}+\hat{\mathcal{V}}^\varphi \boldsymbol{e_{\hat\varphi}}$ are therefore
\begin{eqnarray}
\hat{\mathcal{V}}^{\hat r}&&=\frac{1}{\gamma [1-\nu\sin\psi\sin\alpha]}-\gamma\nu\sin\psi\sin\alpha\nonumber\\
&&=-\gamma\nu^2\left[\frac{1+\sin^2\psi\sin^2\alpha}{1-\nu\sin\psi\sin\alpha}\right],\label{rad1}\\
\hat{\mathcal{V}}^{\hat \theta}&&=-\gamma\nu\cos\psi \label{rad3},\\
\hat{\mathcal{V}}^{\hat\varphi}&&=-\gamma\nu\sin\psi\cos\alpha,\label{rad2}\\
\hat{\mathcal{V}}^{\hat t}&&=\nu(\hat{\mathcal{V}}^{\hat r}\sin\alpha\sin\psi+\hat{\mathcal{V}}^{\hat\theta}\cos\psi+\hat{\mathcal{V}}^{\hat\varphi}\cos\alpha\sin\psi)\nonumber\\
&&=\gamma \nu\left[\frac{\sin\psi\sin\alpha-\nu}{1-\nu\sin\psi\sin\alpha}\right],\label{rad4}
\end{eqnarray}
where the second equality of Eq. (\ref{rad4}) is due to the orthogonality of the $(\boldsymbol{\hat{\mathcal{V}}},\boldsymbol{U})$ pair and we have simplified the components of $\boldsymbol{\hat{\mathcal{V}}}$ of the radiation field. 

\subsubsection{General relativistic equations of motion}
In order to make the equations of motion for the test particle moving in a 3D space explicit, Eqs. (\ref{geom}), we consider the ZAMO frame components of the test particle acceleration $\bold{a}(U)$, Eqs. (\ref{acc1}) -- (\ref{acc4}), and the ZAMO frame components of the radiation force field $\boldsymbol{{\mathcal F}}_{\rm (rad)}(U)$, Eqs. (\ref{enepart}), (\ref{rad1}) -- (\ref{rad4}). The motion of the test particle is completely defined by the following six parameters $(r,\theta,\varphi,\nu,\psi,\alpha)$, the first three describing the position and the last three the velocity field. The displacement field is simply described by $(U^r,U^\theta,U^\varphi)\equiv(dr/d\tau,d\theta/d\tau,d\varphi/d\tau)$. Instead the velocity field is connected to Eqs. (\ref{geom}) for determining $(d\nu/d\tau,d\psi/d\tau,d\alpha/d\tau)$. We note that using Eq. (\ref{acc4}), it is possible to isolate $d\nu/d\tau$, indeed $a(U)^{\hat t}$ is the energy balance equation (see discussions in \cite{Defalco2018}). Then by using the expression of $d\nu/d\tau$ in $a(U)^{\hat\theta}$, Eq. (\ref{acc3}), it is possible to determine $d\psi/d\tau$. Finally using the expressions of $d\nu/d\tau$ and $d\psi/d\tau$ in $a(U)^{\hat r}$, Eq. (\ref{acc1}) yields $d\alpha/d\tau$. 

Therefore, the general relativistic equations in the Kerr metric for the 3D motion of a test particle immersed in the radiation field defined in Sec. \ref{sec:phot} and \ref{sec:eoms} are the following six coupled ordinary differential equations of the first order
\begin{eqnarray}
&&\frac{d\nu}{d\tau}= -\frac{1}{\gamma}\left\{ \sin\alpha \sin\psi\left[a(n)^{\hat r}\right.\right.\label{EoM1}\\
&&\left.\left.\ \quad +2\nu\cos \alpha\sin\psi\, \theta(n)^{\hat r}{}_{\hat \varphi} \right]+\cos\psi\left[a(n)^{\hat \theta}\right. \right.\nonumber\\
&&\left.\left.\ \quad+2\nu\cos\alpha\sin\psi\, \theta(n)^{\hat \theta}{}_{\hat \varphi}\right]\right\}+\frac{\tilde{\sigma}[\Phi E(U)]^2}{\gamma^3\nu}\hat{\mathcal{V}}^{\hat t},\nonumber\\
&&\frac{d\psi}{d\tau}= \frac{\gamma}{\nu} \left\{\sin\psi\left[a(n)^{\hat \theta}+k_{\rm (Lie)}(n)^{\hat \theta}\,\nu^2 \cos^2\alpha\right.\right.\label{EoM2}\\
&&\left.\left.\ \quad+2\nu\cos \alpha\sin^2\psi\ \theta(n)^{\hat \theta}{}_{\hat \varphi}\right]-\sin \alpha\cos\psi \left[a(n)^{\hat r}\right.\right.\nonumber\\
&&\left.\left.\ \quad+k_{\rm (Lie)}(n)^{\hat r}\,\nu^2+2\nu\cos \alpha\sin\psi\, \theta(n)^{\hat r}{}_{\hat \varphi}\right]\right\}\nonumber\\
&&\ \quad+\frac{\tilde{\sigma}[\Phi E(U)]^2}{\gamma\nu^2\sin\psi}\left[\hat{\mathcal{V}}^{\hat t}\cos\psi-\hat{\mathcal{V}}^{\hat \theta}\nu\right],\nonumber\\
&&\frac{d\alpha}{d\tau}=-\frac{\gamma\cos\alpha}{\nu\sin\psi}\left[a(n)^{\hat r}+2\theta(n)^{\hat r}{}_{\hat \varphi}\ \nu\cos\alpha\sin\psi\right.\label{EoM3}\\
&&\left.\ \quad+k_{\rm (Lie)}(n)^{\hat r}\,\nu^2+k_{\rm (Lie)}(n)^{\hat \theta}\,\nu^2\cos^2\psi \sin\alpha\right]\nonumber\\
&&\ \quad+\frac{\tilde{\sigma}[\Phi E(U)]^2\cos\alpha}{\gamma\nu\sin\psi}\left[\hat{\mathcal{V}}^{\hat r}-\hat{\mathcal{V}}^{\hat \varphi}\tan\alpha\right],\nonumber\\
&&U^r\equiv\frac{dr}{d\tau}=\frac{\gamma\nu\sin\alpha\sin\psi}{\sqrt{g_{rr}}}, \label{EoM4}\\
&&U^\theta\equiv\frac{d\theta}{d\tau}=\frac{\gamma\nu\cos\psi}{\sqrt{g_{\theta\theta}}} \label{EoM5},\\
&&U^\varphi\equiv\frac{d\varphi}{d\tau}=\frac{\gamma\nu\cos\alpha\sin\psi}{\sqrt{g_{\varphi\varphi}}}-\frac{\gamma N^\varphi}{N},\label{EoM6}
\end{eqnarray}
where  $\tilde{\sigma}=\sigma/m$ and the two angles $\beta$ and $\zeta$ are calculated in terms of the two impact parameters $b$ and $q$. For $\psi=\zeta=\pi/2$ the  equations of motion reduce to the 2D case \cite{Bini2009}. Such set of equations reduce also to the classical 3D case in the weak field limit (see Appendix \ref{sec:classicPR}).

Following \cite{Abramowicz1990,Bini2009,Bini2011} we define the relative luminosity of the radiation field as 
\begin{equation}\label{eq:A}
A=\tilde{\sigma}\Phi_0^2E^2\,.
\end{equation}
Eq. (\ref{eq:A}) can be recast in the terms of the relative luminosity $A=L_\infty/L_{\rm EDD}$, taking thus the values in $[0,1]$, where $L_\infty$ is the luminosity of the central source as seen by an observer at infinity and $L_{\rm EDD}=4\pi Mm/\sigma$ is the Eddington luminosity at infinity. Then for the investigated radiation field with zero angular momenta ($b=0$, $\beta=\pi/2$) and without latitudinal photon motion ($q=- a^{2} \cos^{2}\theta$, $\zeta=\pi/2$), the term $\tilde{\sigma}[\Phi E(U)]^2$ becomes 
\begin{equation}
\tilde{\sigma}[\Phi E(U)]^2=\frac{ A\,\gamma^2\, [1-\nu\sin\psi\sin\alpha]^2}{N^2 \sqrt{(r^2+a^2)^2-a^2\,\Delta\,\sin^{2}\theta}} \,.
\end{equation}

\section{Critical hypersurface}
\label{sec:critc_rad}
The system of six differential equations (\ref{EoM1}) -- (\ref{EoM6}) admits a critical solution of radial equilibrium, which corresponds to the axially-symmetric hypersurface where radiation pressure balances the attraction of the gravitational field. 
Let us consider a test particle moving purely radially with respect to the ZAMO frame ($\alpha=\psi=\pm\pi/2$). Then, at the critical radius $r_{\rm (crit)}$, where the test particle is in rest with respect to the ZAMO frame ($\nu=0, \gamma=1$), the first equation of motion, Eq. (\ref{EoM1}), takes the form
\begin{equation} \label{r-equilibrium}
a(n)^{\hat r}=\frac{A}{N^2 \sqrt{(r_{\rm (crit)}^2+a^2)^2-a^2\,\Delta_{\rm (crit)}\,\sin^{2}\theta}}. 
\end{equation}
In the case of pure radial motion ($\cos\alpha=0, \frac{d\alpha}{d\tau}=0 $), the third equation of motion, Eq. (\ref{EoM3}), is automatically fulfilled. If we multiply the second equation of motion, Eq. (\ref{EoM2}), by the term $\nu^2$, thus removing its divergence, one can easily see that it is fulfilled in the radial equilibrium case ($\cos\psi=0, \frac{d\psi}{d\tau}=0,\nu=0 $). 
For $\theta=\pi/2$ (i.e. in the equatorial plane) relation (\ref{r-equilibrium}) corresponds to  the equilibrium condition Eq. (2.33) derived in \cite{Bini2009}, which gives the values $r_{\rm(crit)}$ of the radial coordinate where the test particle comoves with the ZAMOs in the equatorial circular orbit. However, relation (\ref{r-equilibrium}) generalizes this condition also for the case of test particles with arbitrary polar angle $\theta$ and therefore describes a critical hypersurface which envelops the central compact object and where the test particles comoves with the local ZAMOs in a bound quasi-circular orbits. \footnote{A different mechanism that leads to the formation of similar off-equatorial circular orbits is the interaction of charged test particles with the magnetic field of a neutron star \cite{Kovar2008}.}
 
\begin{figure*}[th!]
\centering
\hbox{
\includegraphics[scale=0.4]{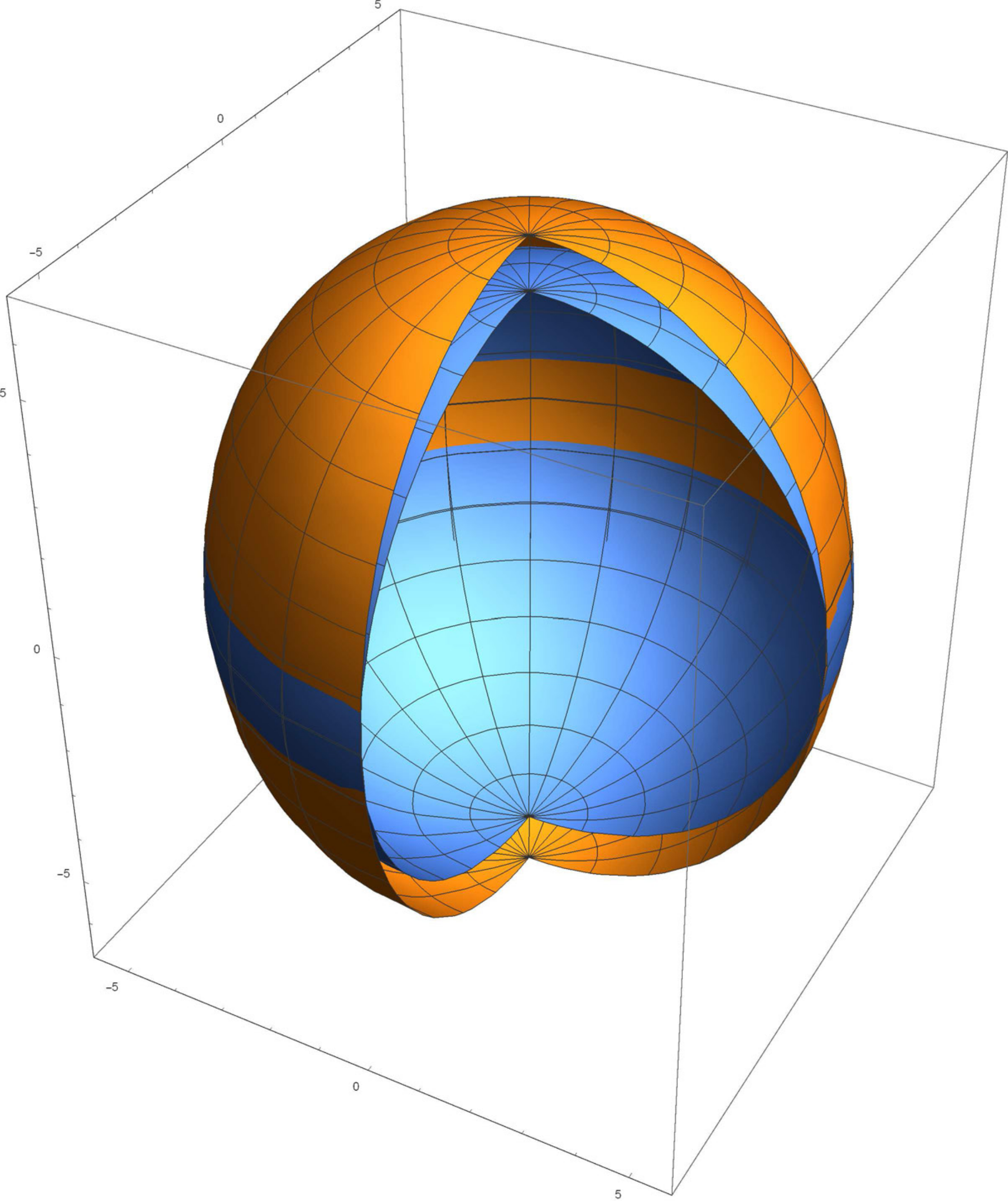}
\hspace{0.5cm}
\includegraphics[scale=0.3]{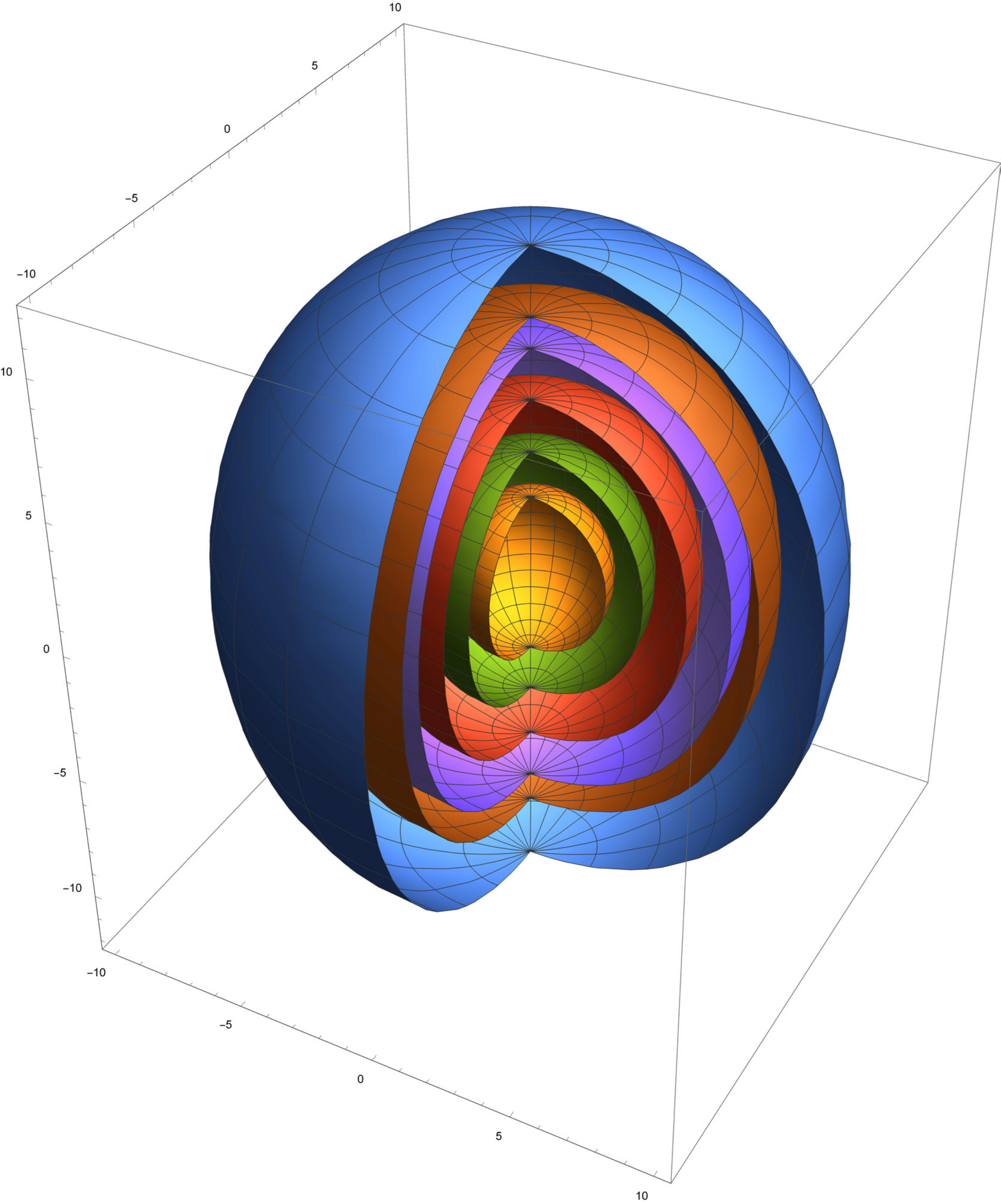}}
\caption{Left panel: Critical hypersurfaces for the case of high spin $a=0.9995$ (orange) and the case of the Schwarzschild spacetime with $a=0$ (blue). For the Schwarzschild case the critical radius is $r_{\rm (crit)} \sim 5.56M$, while for the Kerr case in the equatorial plane is $r^{\rm eq}_{\rm(crit)} \sim 5.52M$ and $r^{\rm pole}_{\rm(crit)} \sim 6.56M$ at the poles. The relative luminosity of the radiating field takes the value of $A=0.8$. Right panel: Critical hypersurfaces for the values of the relative luminosity $A=0.5, \,0.7, \,0.8, \,0.85, \,0.87, \,0.9$ at a constant spin $a=0.9995$. The respective critical radii in the equatorial plane are $r^{\rm eq}_{\rm(crit)} \sim 2.71M,  4.01M, 5.52M, 7.04M, 7.99M, 10.16M$, while at poles they are $r^{\rm pole}_{\rm(crit)} \sim 2.97M,  4.65M,  6.56M,  8.38M,  9.48M, 11.9M$.}
\label{fig:Fig2}
\end{figure*}

 In the case of a non-zero spin, the critical radius given by Eq. (\ref{r-equilibrium}) is function of the polar angle of $r_{\rm(crit)}=r_{\rm (crit)}(A,\theta)$ (in addition to the relative luminosity $A$). The radial equilibrium therefore occurs at the axially symmetric hypersurface, whose shorter axis lies in the equatorial plane and longer axis in the polar direction.
This is due to the properties of frame-dragging, as photons (and test particles) are dragged maximally in the azimuthal direction in the equatorial plane $\theta=\pi/2$. Therefore the radial component of the photon four-velocity reaches a maximum (and thus the radial momentum transfer is largest) along the polar axis and decreases for increasing polar angles; that is the reason why the critical hypersurface is elongated along the polar axis. In the case of zero spin (Schwarzschild spacetime), the critical hypersurface turns into a sphere with radius corresponding to the value given by Eq. (2.33) in \cite{Bini2009}. The left panel of Fig. \ref{fig:Fig2} compares the shape of the critical hypersurfaces for a high-spin Kerr spacetime $a=0.9995$ and for a Schwarzschild spacetime with $a=0$, where the relative luminosity of the radiating field is in both cases set to the value of $A=0.8$. In the high-spin case the critical radius is $r^{\rm eq}_{\rm(crit)} \sim 5.52M$ in the equatorial plane and $r^{\rm pole}_{\rm(crit)} \sim 6.56M$ at the poles. In the case of a Schwarzschild spacetime, the radius of the critical sphere is $r_{\rm (crit)} \sim 5.56M$. The right panel of Fig. \ref{fig:Fig2} illustrates the shape of the critical hypersurfaces for the values of the relative luminosity in the interval $0.5-0.9$ and for a constant value of the spin $a=0.9995$. 

\section{Test particle orbits}
\label{sec:orbits}
We have developed the  \emph{3D PRtrajectories} code to integrate the test particles trajectories described by equations (\ref{EoM1}) -- (\ref{EoM6}). The integration of the equations of motion in three spatial dimensions turns out to be substantially more sensitive to integration errors than the the 2D case. Therefore we adapted the highly-accurate core for the integration of photon trajectories used in  \emph{LSDCode+}  \cite{Bakala2015} to the case of massive particles. The code
implements the Runge-Kutta method of the eighth order (the Dorman -- Prince method) \cite{Press2002} with an adaptive step. Successful integration of the 3D trajectory of test particles influenced by the radiation field (especially in the latitudinal direction) requires advanced monitoring of integration errors. In the  \emph{3D PRtrajectories} code, the  \emph{PI stepsize control} algorithm (see \cite{Press2002} for details) is implemented, which easily attains  an average relative accuracy of $\sim 10^{-14}$. Such a value allow precise and consistent integration of 3D trajectories even in the most sensitive parts, the vicinity of turning points.

We integrated equations (\ref{EoM1}) -- (\ref{EoM6}) for a set of different boundary conditions and model parameters.  
Our results show that the main qualitative features of the 2D case examined in \cite{Bini2009,Bini2011} remain the same for the trajectories in three spatial dimensions. Similarly to the 2D case, we can divide the orbits into two distinct classes depending on the initial radial position $r(0)=r_0$: inside and outside the critical hypersurface. Also in the 3D case, a test particle trajectory can have only two possible ends, either $(i)$ it goes to infinity or $(ii)$ it reaches the critical hypersurface. Moreover in the presence of (an outgoing) radiation field, the test particle cannot cross the event horizon. We compared representative trajectories of test particles with a polarly and azimuthally-oriented initial velocity for the case of a Schwarzschild spacetime, for the case of the Kerr metrics with a  small spin (a=0.05) that approximates the spacetime in the vicinity of NSs and, finally, for the case of the Kerr metrics with very high spin ($a=0.9995$) that corresponds to the spacetime in the vicinity of almost extreme BHs.
 
\begin{figure*}[th!]
\centering
\hbox{
\includegraphics[scale=0.28]{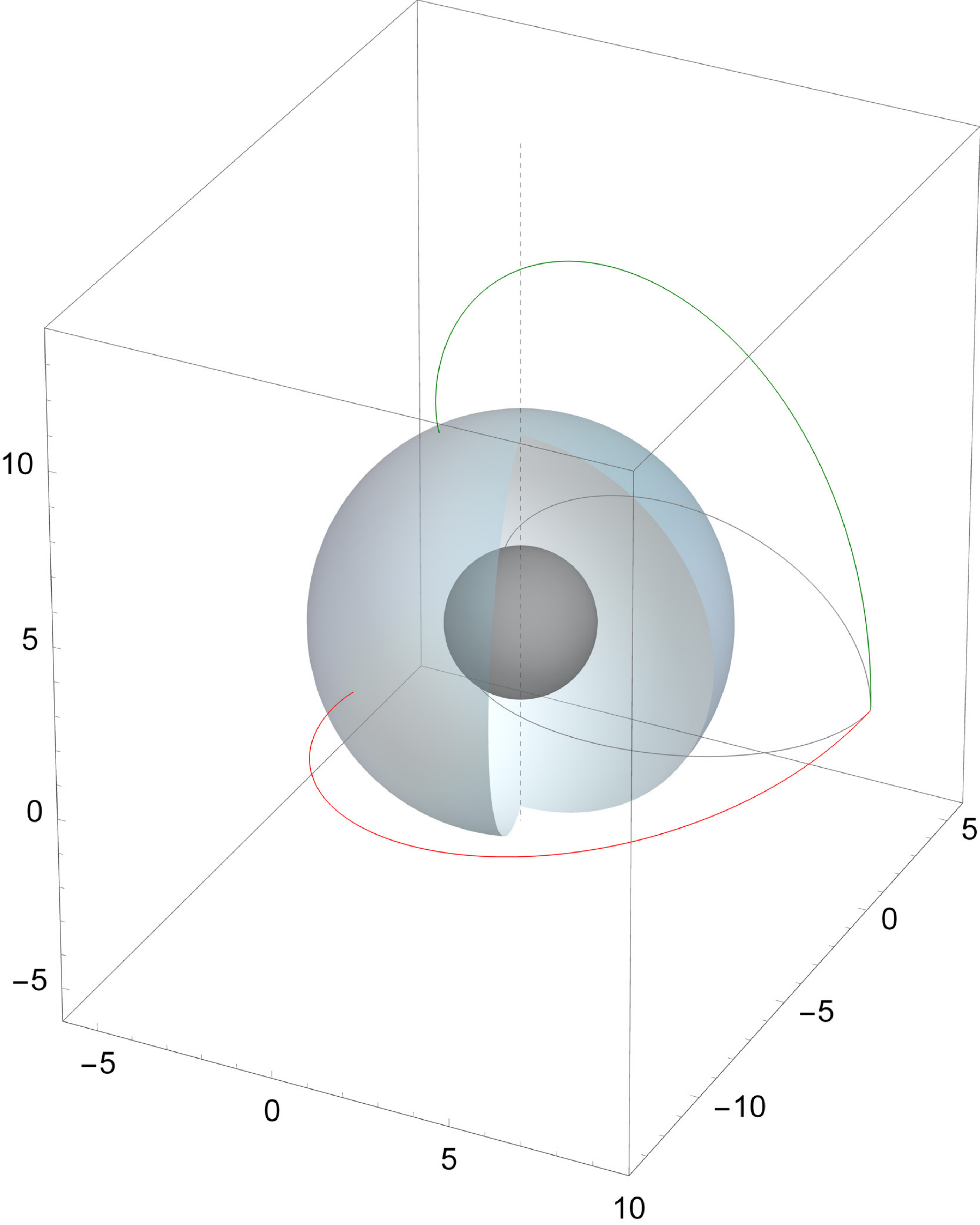}
\hspace{0.5cm}
\includegraphics[scale=0.3]{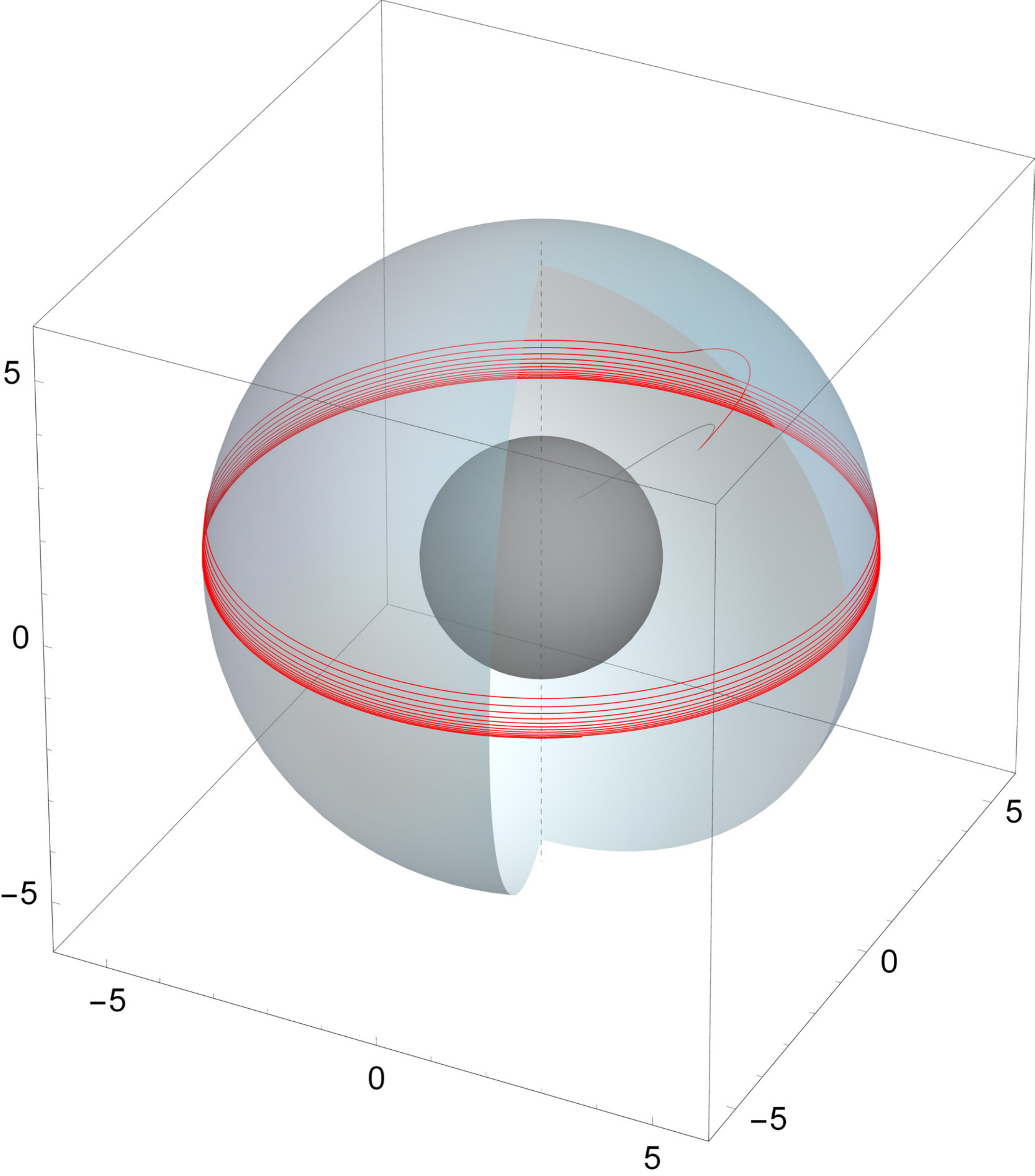}}
\caption{Left panel: test particle trajectories in a Schwarzschild geometry under the influence of a radiation field  with $A=0.8$. Test particles start  at $r_0=8M$ in the equatorial plane with  initial velocity $\nu_0=0.8$ in azimuthal (red) and polar (green) directions. Right panel: test particle trajectories in a Kerr geometry with small spin ($a=0.05$) under the influence of a radiation field with $A=0.8$. A test particle starts inside the critical hypersurface at $r_0=4M,\, \theta_0=\pi/4$ with initial velocity $\nu_0=0.4$ in the azimuthal direction. In both panels the inner dark surface represents the event horizon and blue-gray, partially open surface represents the critical hypersurface. Gray curves show the geodesic trajectories (i.e. $A=0$) for test particles with initial conditions equal to those described above.}
\label{fig:Figs_S}
\end{figure*}

In the case of the Schwarzschild metric our results fully agree with those from earlier analyses of the 2D case in which motions are confined to the equatorial plane \cite{Bini2009,Bini2011} (note however, that this can be chosen arbitrarily for spherically symmetric metrics and radiation fields). The left panel of Fig. \ref{fig:Figs_S} shows examples of 3D trajectories which reflect the spherically-symmetric limit of equations (\ref{EoM1} - \ref{EoM6}) in the case of zero spin. The trajectories of test particles starting from the same location but with initial velocity oriented polarly and azimuthally are identical except for the different orientation of the plane on which they lie. In such a case, the 3D trajectories are easily transformed to the corresponding 2D trajectories through coordinate rotation. The left panel of Fig.\ref{fig:Figs_S} then shows that in a Schwarzschild spacetime, once the trajectory of a test particle  reaches the spherical critical hypersurface it stops precisely there.

\begin{figure*}[th!]
\centering
\hbox{
\includegraphics[scale=0.28]{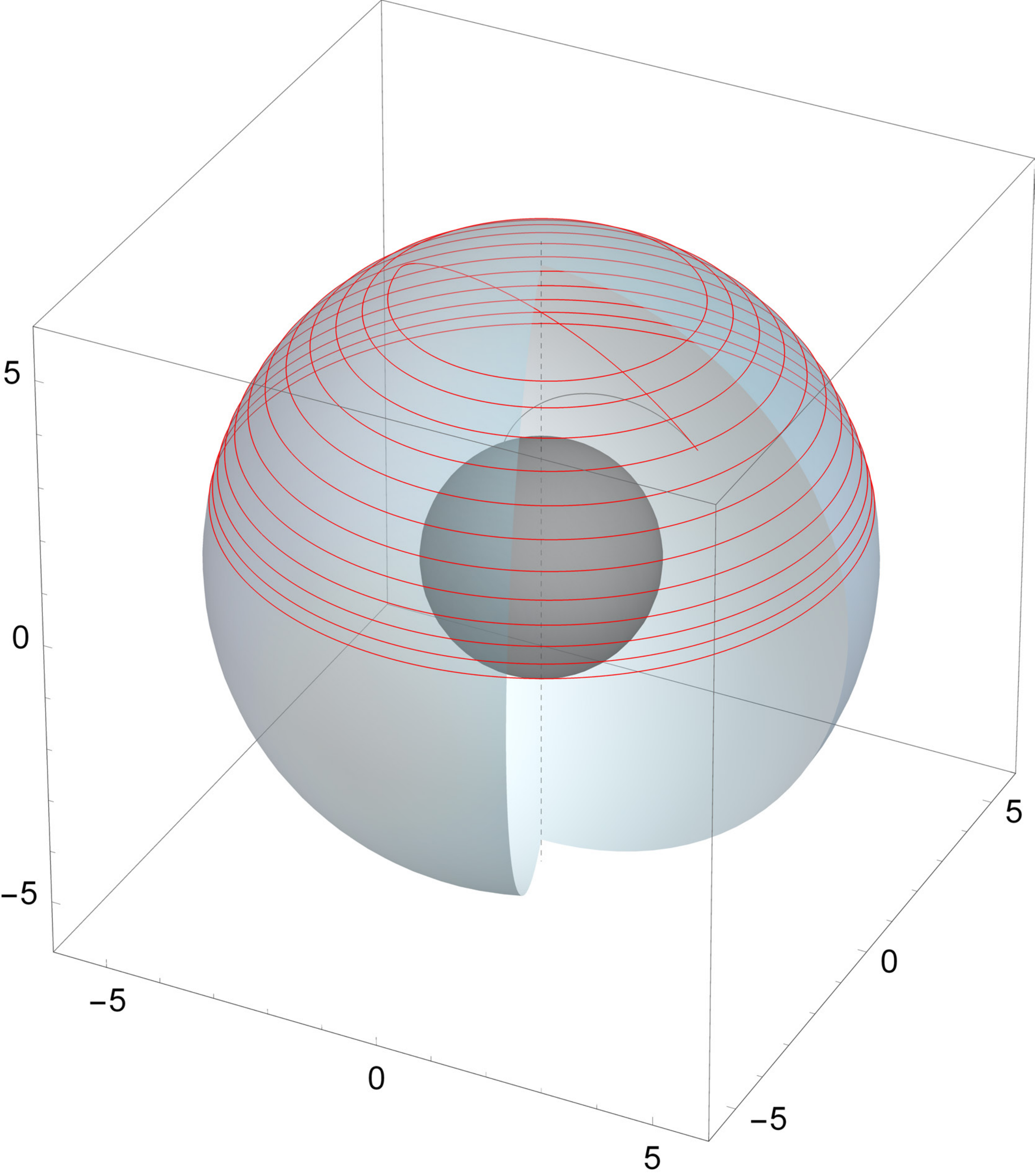}
\includegraphics[scale=0.32]{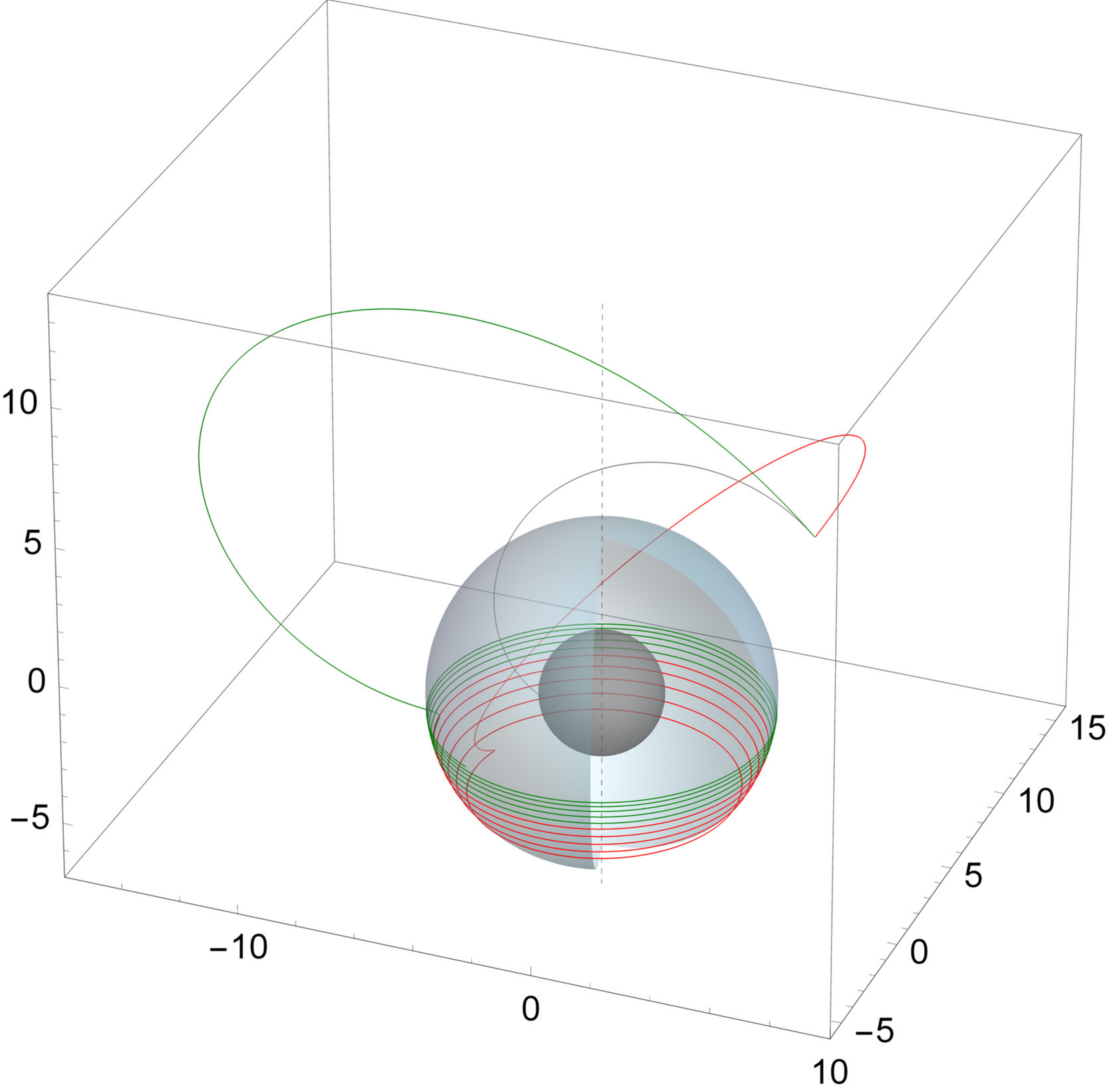}}
\caption{Test particle trajectories in a Kerr geometry with small spin ($a=0.05$) under the influence of a radiation field with $A=0.8$. Left panel: the test particle starts its motion inside the critical hypersurface at $r_0=4M,\, \theta_0=\pi/4$ with initial velocity $\nu_0=0.4$ in the polar direction. The gray curve denotes the geodesic trajectory (i.e. $A=0$) with $\nu_0=0.4$. Right panel: the test particles start their motion outside the critical hypersurface at $r_0=10M,\, \theta_0=\pi/4$ with initial velocity $\nu_0=0.4$ in the azimuthal (red curve) and polar direction (green curve). In both panels the inner dark surface represents the event horizon and blue-gray, partially open surface represent the critical hypersurface.}
\label{fig:Figs_KSI1}
\end{figure*}

\begin{figure*}[th!]
\centering
\hbox{
\includegraphics[scale=0.28]{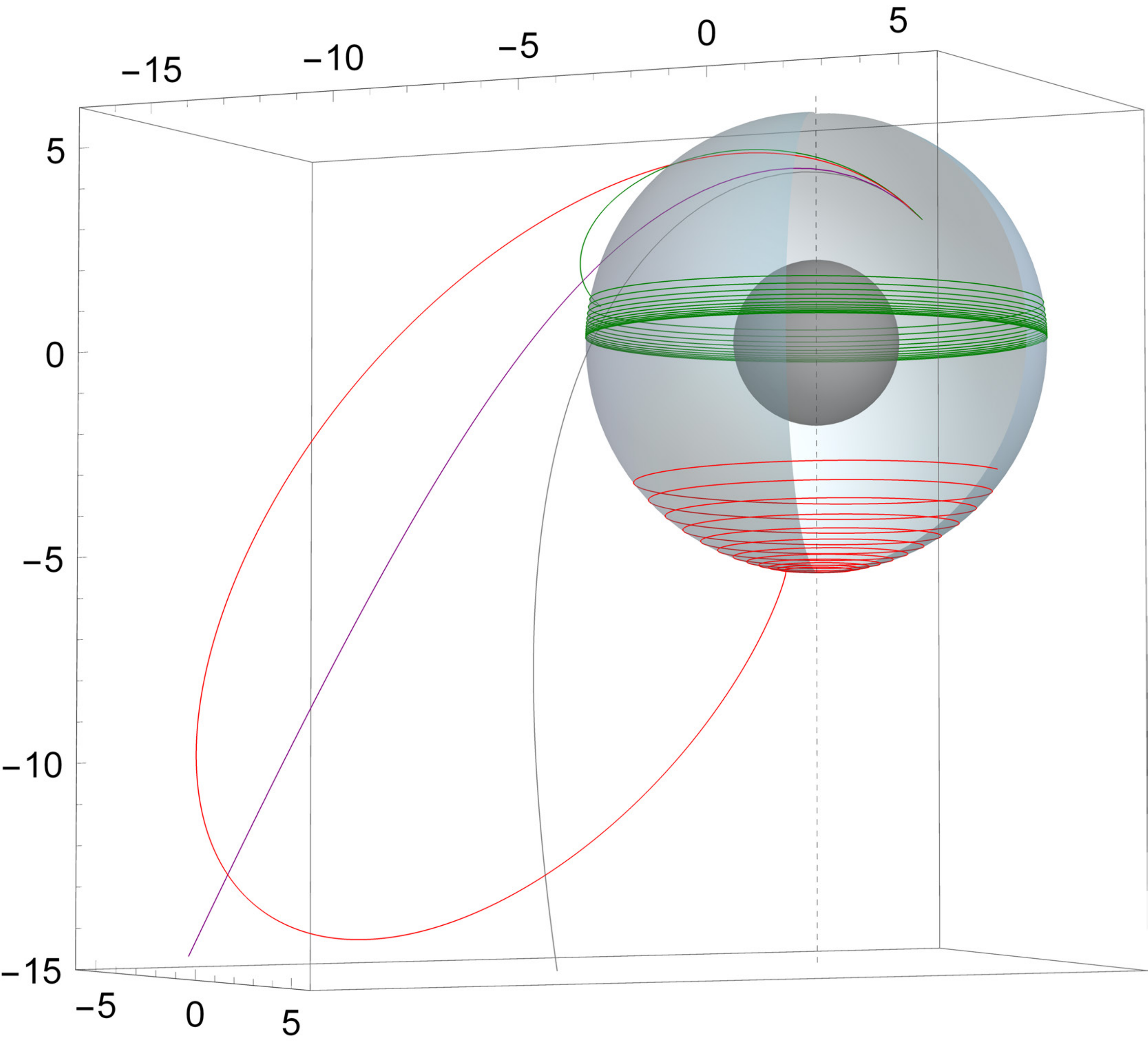}
\includegraphics[trim=1cm 1cm 0cm 1cm, scale=0.38]{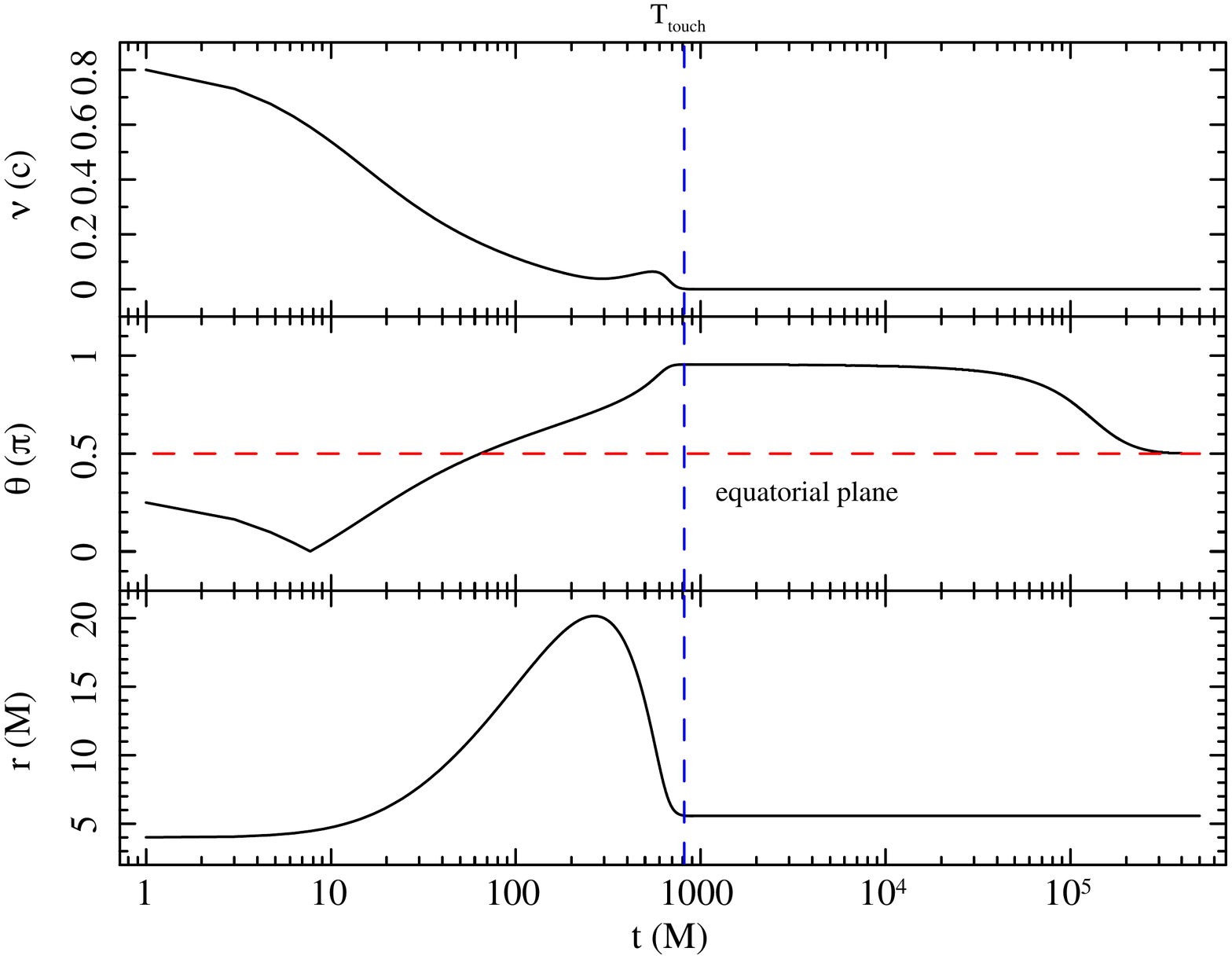}}
\caption{Test particle trajectories in a Kerr geometry with small spin ($a=0.05$) under the influence of a radiation field with $A=0.8$. Left panel: three test particle starting their motion outside the critical hypersurface at $r_0=4M,\, \theta_0=\pi/4$ with initial velocity along the polar direction and values $\nu_0=0.6$ (green curve), $\nu_0=0.8$ (red curve), $\nu_0=0.87$ (violet curve - escape trajectory). The gray curve shows corresponding the geodesic trajectory (i.e. $A=0$) with initial velocity $\nu_0=0.8$. The inner dark surface represent the event horizon and blue-gray, partially open (spherical or quasi-spherical) surface represents the critical hypersurface. Right panel: velocity profile $\nu$, latitudinal angle $\theta$, and radius $r$ in terms of coordinate time $t$ for the test particle motion with $\nu_0=0.8$ (red curve in left panel). The vertical dashed blue line, $T_{\rm touch}$, represents the time at which the test particle reaches the critical hypersurface; from there on the latitudinal drift on the hypersurface sets in (note the velocity in this stage in much lower  than velocities off the hypersurface). The horizontal dashed red line represents the equatorial plane.}
\label{fig:Figs_KSI2}
\end{figure*}

\begin{figure*}[th!]
\centering
\hbox{
\includegraphics[width=4.4cm, height=11cm]{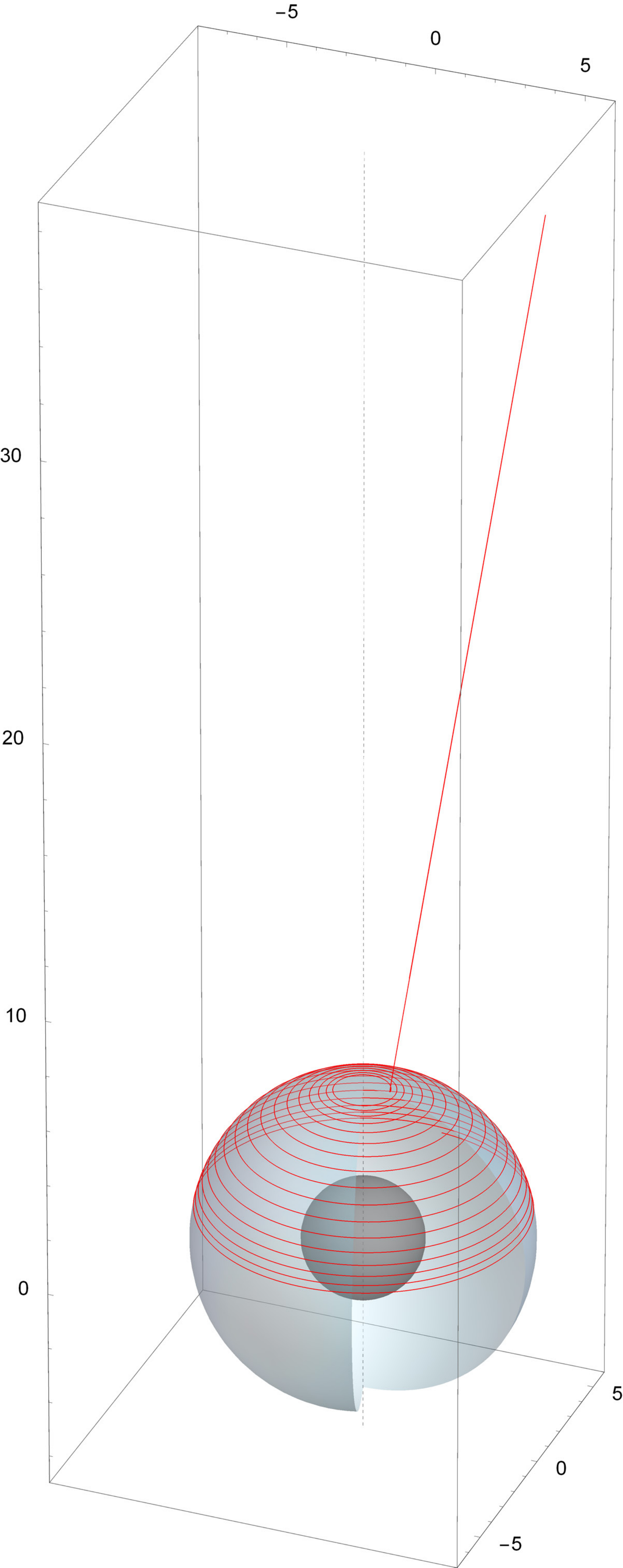}
\includegraphics[trim=1cm 1cm 0cm 1cm, scale=0.55]{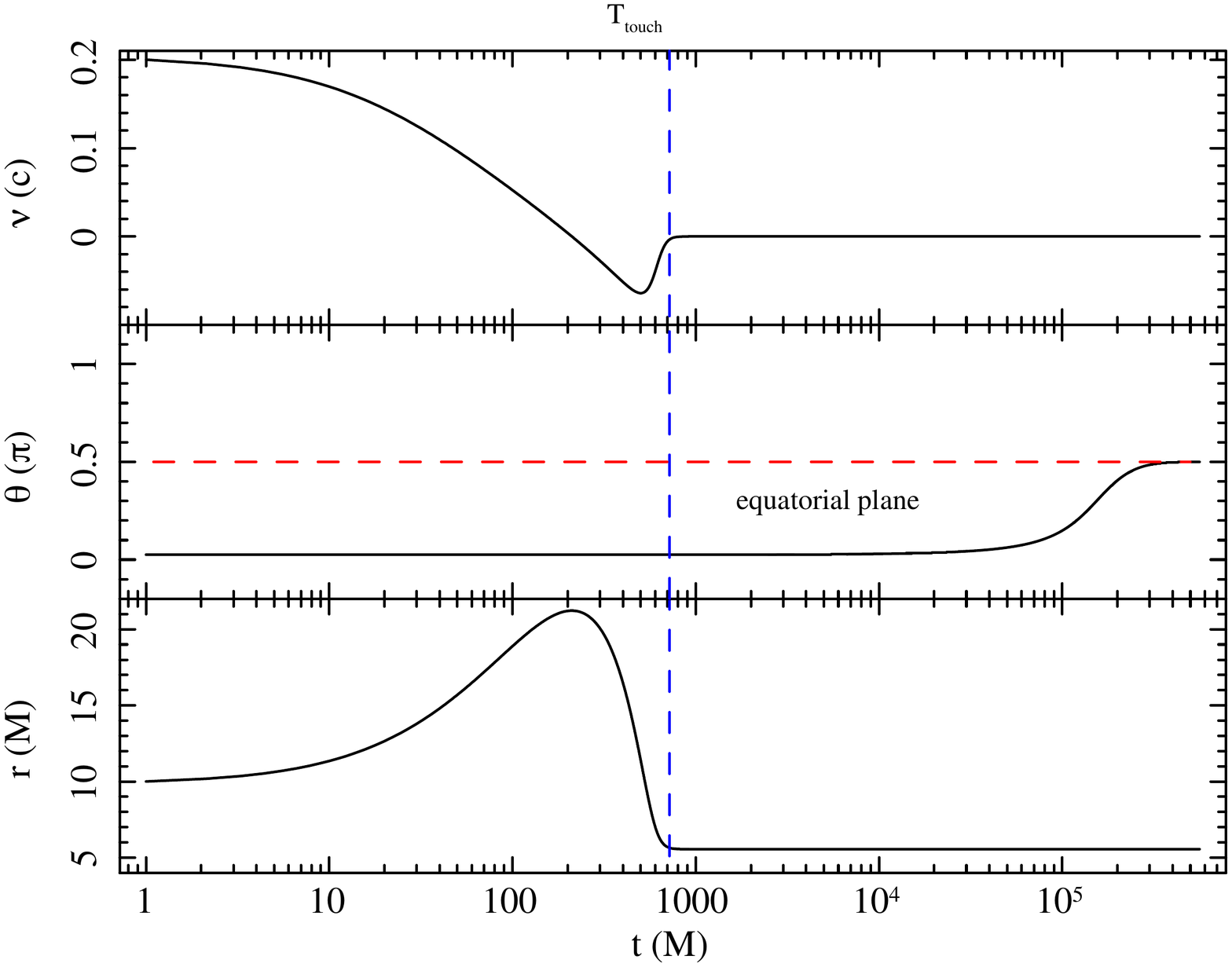}}
\caption{Test particle trajectories in a Kerr geometry with small spin ($a=0.05$) under the influence of a radiation field with $A=0.8$. Left panel: the test particle starts its motion outside the critical hypersurface at $r_0=10M,\, \theta_0=\pi/20$ with initial velocity $\nu_0=0.4$. The inner dark surface represent the event horizon and blue-gray, partially open (spherical or quasi-spherical) surface represents the critical hypersurface. Right panel: velocity profile $\nu$, latitudinal angle $\theta$, and radius $r$ in terms of coordinate time $t$ for the test particle motion with $\nu_0=0.4$ (red curve in left panel). The vertical dashed blue line, $T_{\rm touch}$, represents the time at which the test particle reaches the critical hypersurface; from there on the latitudinal drift on the hypersurface sets in (note the velocity in this stage in much lower  than velocities off the hypersurface). The horizontal dashed red line represents the equatorial plane.}
\label{fig:Figs_KSO2}
\end{figure*}

\begin{figure*}[th!]
\centering
\hbox{
\includegraphics[width=8cm, height=8cm]{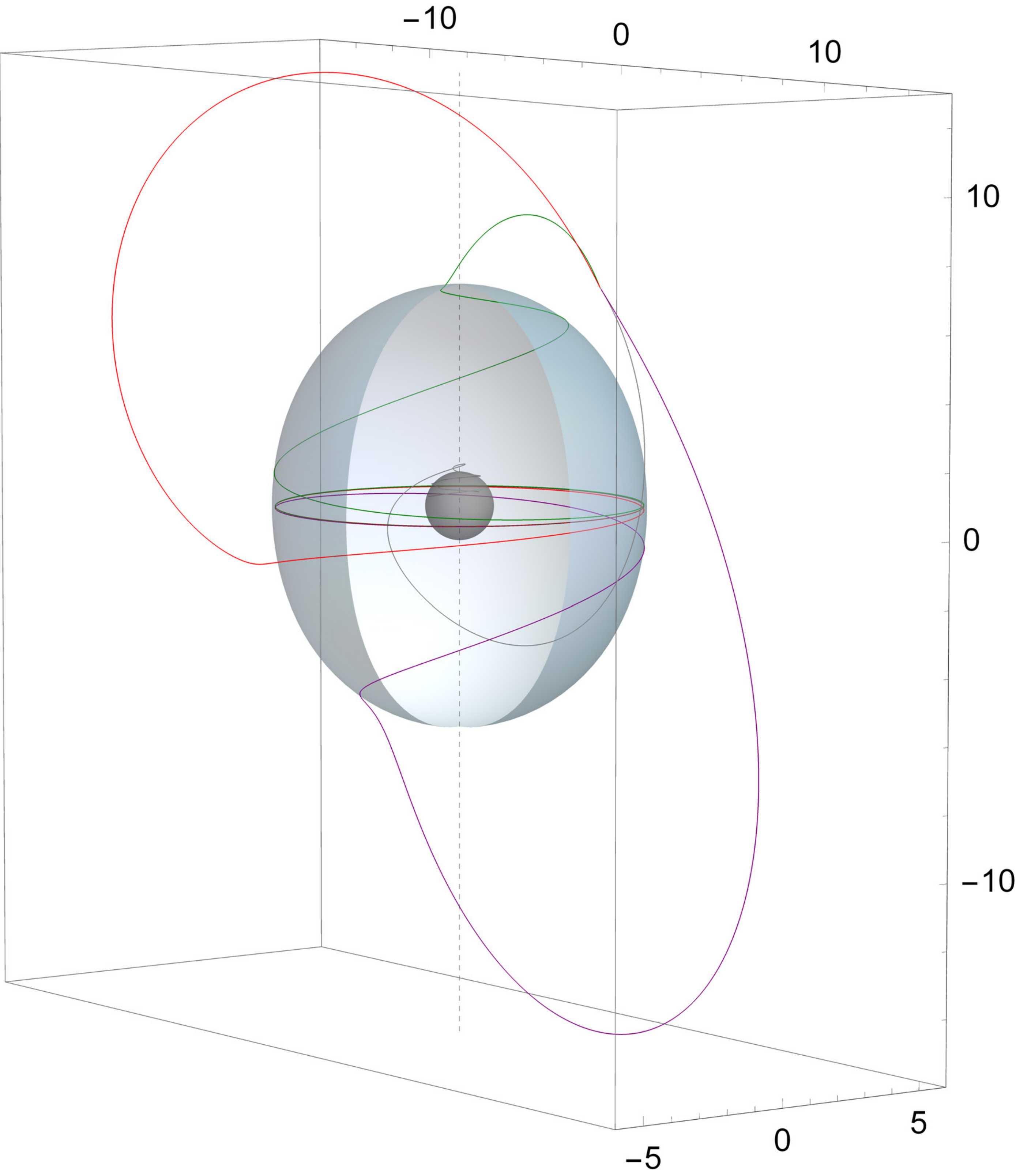}
\includegraphics[trim=1cm 1cm 0cm 1cm, scale=0.4]{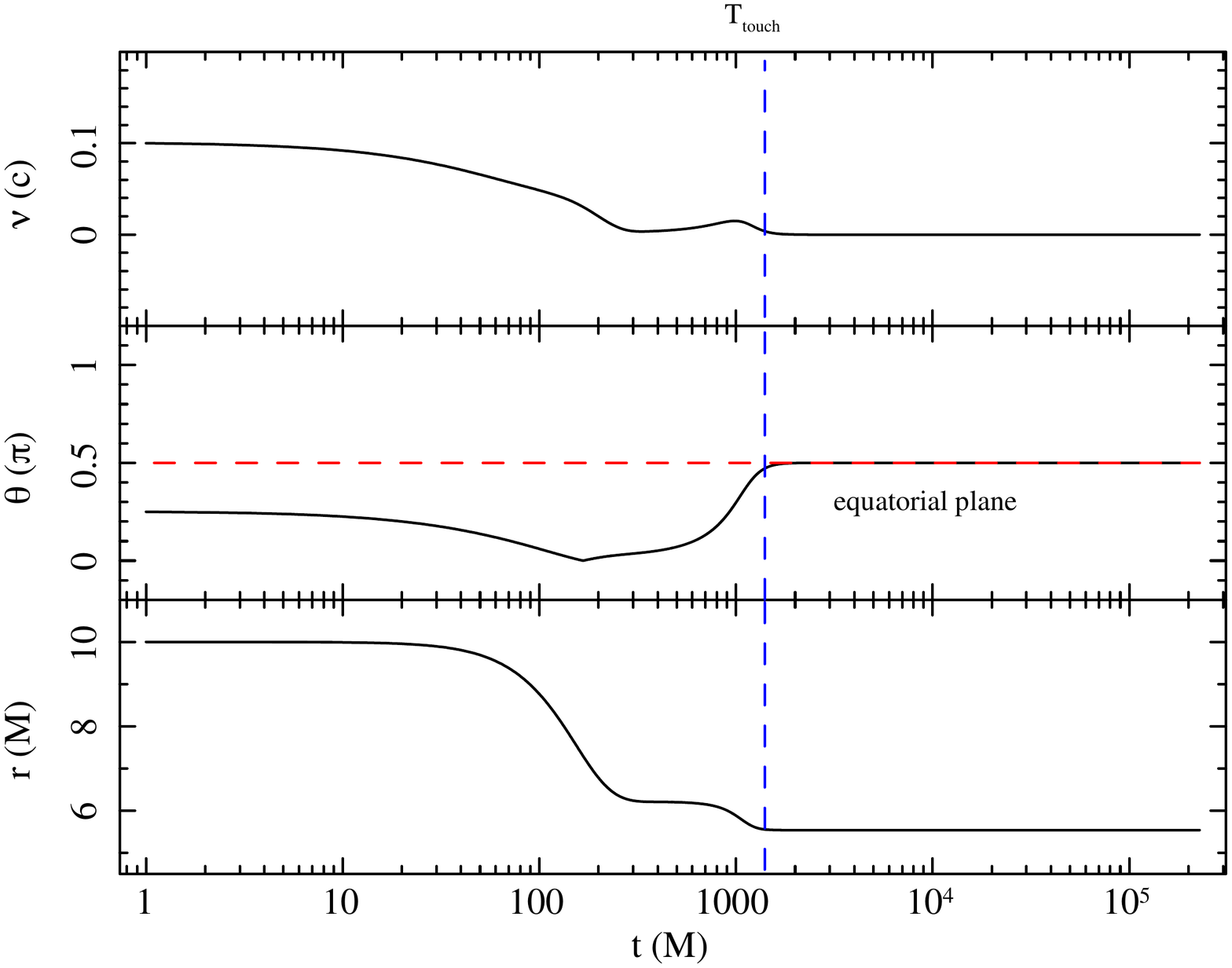}}
\caption{Test particle trajectories in a Kerr geometry with almost extreme spin ($a=0.9995$) under the influence of a radiation field with $A=0.8$. Left panel: three test particle starting their motion outside the critical hypersurface at $r_0=10M,\, \theta_0=\pi/4$ with initial velocity along the polar direction and values oriented towards the north pole $\nu_0=0.1$ (green curve), $\nu_0=0.25$ (red curve), and oriented towards the south pole $\nu_0=0.25$ (violet curve). The inner dark surface represent the event horizon and blue-gray, partially open (spherical or quasi-spherical) surface represents the critical hypersurface. Right panel: velocity profile $\nu$, latitudinal angle $\theta$, and radius $r$ in terms of coordinate time $t$ for the test particle motion with $\nu_0=0.1$ (green curve in left panel). The vertical dashed blue line, $T_{\rm touch}$, represents the time at which the test particle reaches the critical hypersurface; from there on the latitudinal drift on the hypersurface sets in (note the velocity in this stage in much lower  than velocities off the hypersurface). The horizontal dashed red line represents the equatorial plane.}
\label{fig:Figs_KH1}
\end{figure*}

\begin{figure*}[th!]
\centering
\hbox{
\includegraphics[width=0.52\textwidth]{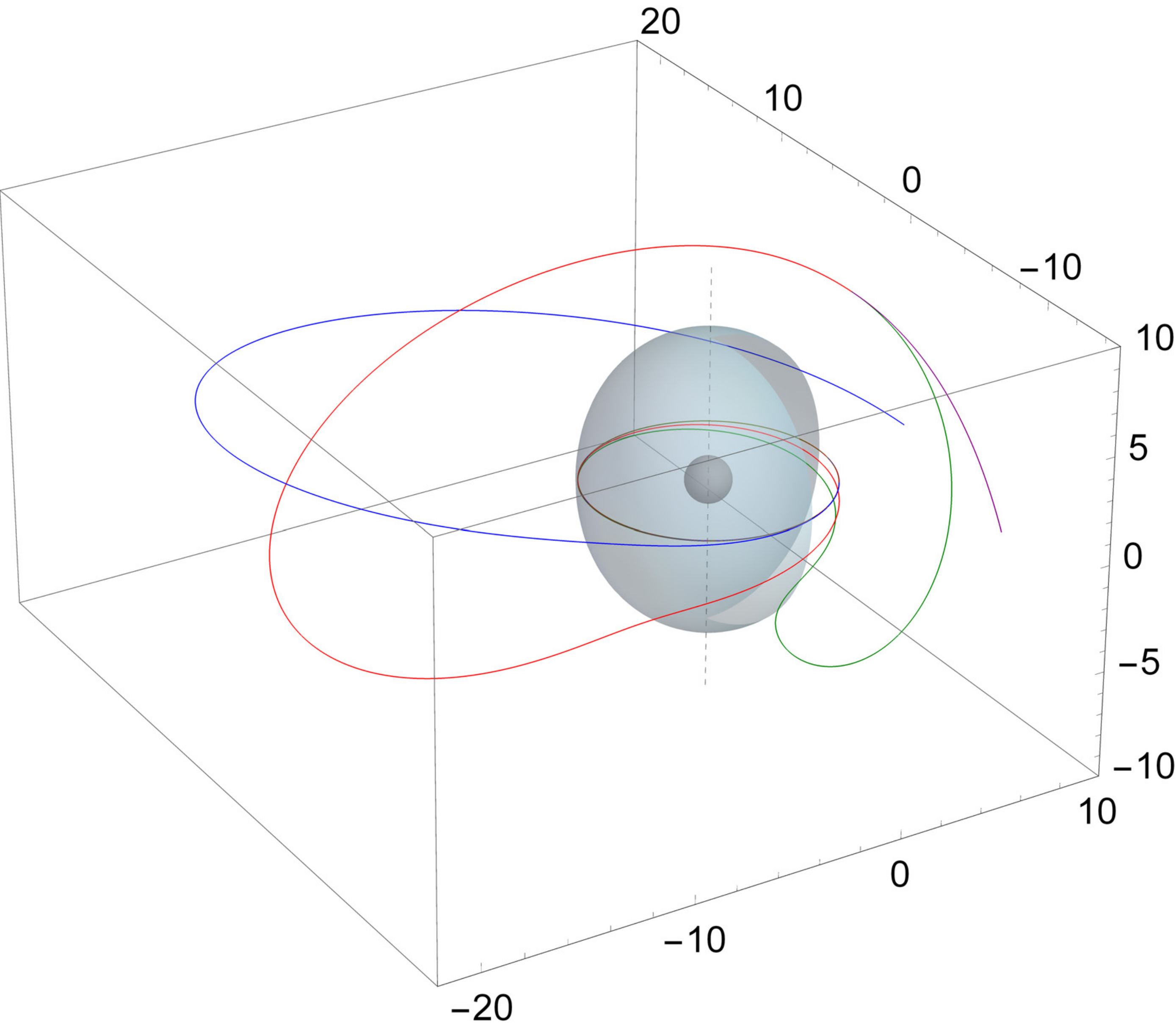}
\hspace{0.5cm}
\includegraphics[width=0.43\textwidth]{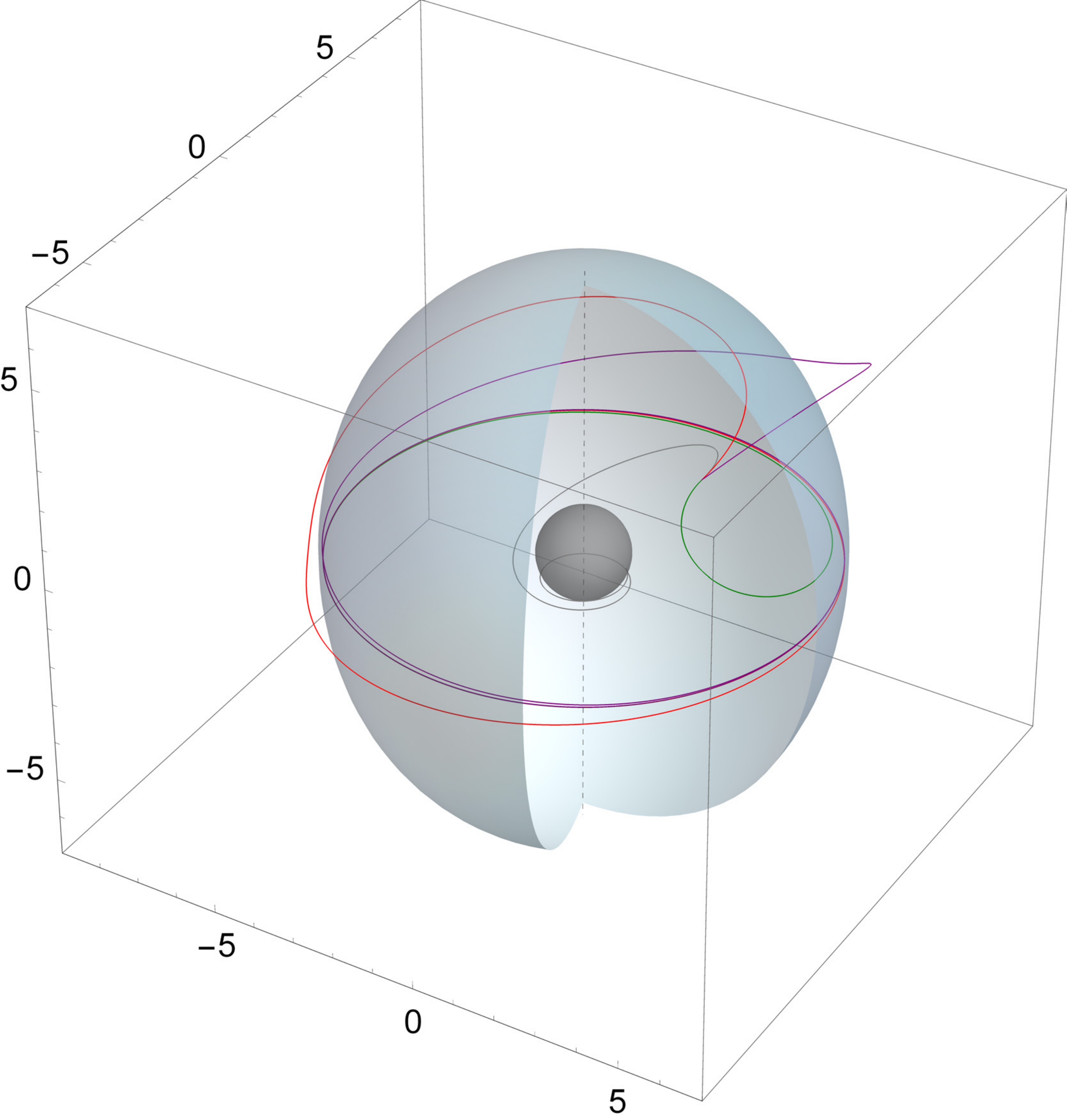}}
\caption{Test particle trajectories in a Kerr geometry with almost extreme spin ($a=0.9995$) under the influence of a radiation field with $A=0.8$. Left panel: three test particle starting their motion outside the critical hypersurface at $r_0=10M$ with initial velocity $\nu_0=0.25$, off-equatorial initial position at $\theta_0=\pi/4$, and in the azimuthal direction co-rotating (red curve) and counter-rotating (green curve) with respect to the compact object and on the equatorial ($\theta_0=\pi/2$) in the azimuthal direction co-rotating with respect to the compact object (blue curve). Right panel: three test particle starting their motion inside the critical hypersurface at $r_0=4M,\,\theta_0=\pi/4$ with initial velocity $\nu_0=0.4$ in the azimuthal direction co-rotating (red curve) and counter-rotating (green curve) with respect to the compact object and in the outgoing radial azimuthal direction (violet curve). The gray curve shows corresponding the geodesic trajectory (i.e. $A=0$) with respect to the red curve. In both panels, the inner dark surface represent the event horizon and blue-gray, partially open (spherical or quasi-spherical) surface represents the critical hypersurface. 
}
\label{fig:Figs_KH2}
\end{figure*}

Note that the presence of even a very small spin ($a=0.05$) breaks the spherical symmetry of the spacetime geometry and radiation field and introduces qualitatively-new features in test particle trajectories, owing to frame dragging effects. In the Kerr case, the test particle, once captured on the critical hypersurface, gets dragged azimuthally at the angular velocity $\Omega_{\mathrm{ZAMO}}$ and undergoes a \emph{latitudinal drift towards the equatorial plane} (see Sec. \ref{sec:latdrift}, for a detailed explanation). Hence the test particle spirals on the critical hypersurface as it shifts to lower and lower latitudes. The results of our numerical integrations show that, besides the angular velocity $\Omega_{\mathrm{ZAMO}}$, the velocity of the latitudinal drift increases for increasing spins. In fact in for small spin values, test particles caught on the critical hypersurface encircle multiple spirals before attaining the final purely circular equatorial trajectory. Such behavior is exhibited by test particles whose motion starts from the inside (see right panel of Fig. \ref{fig:Figs_S} and left panel of Fig. \ref{fig:Figs_KSI1}), as well as the outside of the critical hypersurface (see Fig. \ref{fig:Figs_KSI2} and right panel of Fig. \ref{fig:Figs_KSI1}). The behaviour of a test particles beginning its motion near the polar axis in a outgoing, purely radial direction is illustrated in Fig. \ref{fig:Figs_KSO2}: the test particle initially travels outward, reaches the turning point and then falls back along a nearly identical trajectory; after being captured near the pole of the critical hypersurface, it drifts toward the equator in tight spiralling trajectory that spans over most of northern hemisphere of the critical hypersurface.
For a value of the spin ($a=0.9995$) close to that of an extreme Kerr BH, frame dragging is faster and  leads to a faster latitudinal drift, besides a higher $\Omega_{\mathrm{ZAMO}}$. Therefore test particles captured on the critical hypersurface at any value of the polar coordinate $\theta$ are dragged quickly to the equatorial plane where they attain a purely circular trajectory (see  Figs \ref{fig:Figs_KH1} and \ref{fig:Figs_KH2}). 


\subsection{Orbits bound to the critical hypersurface}
\label{sec:latdrift}
\begin{figure*}[th!]
\centering
\hbox{
\includegraphics[scale=0.3]{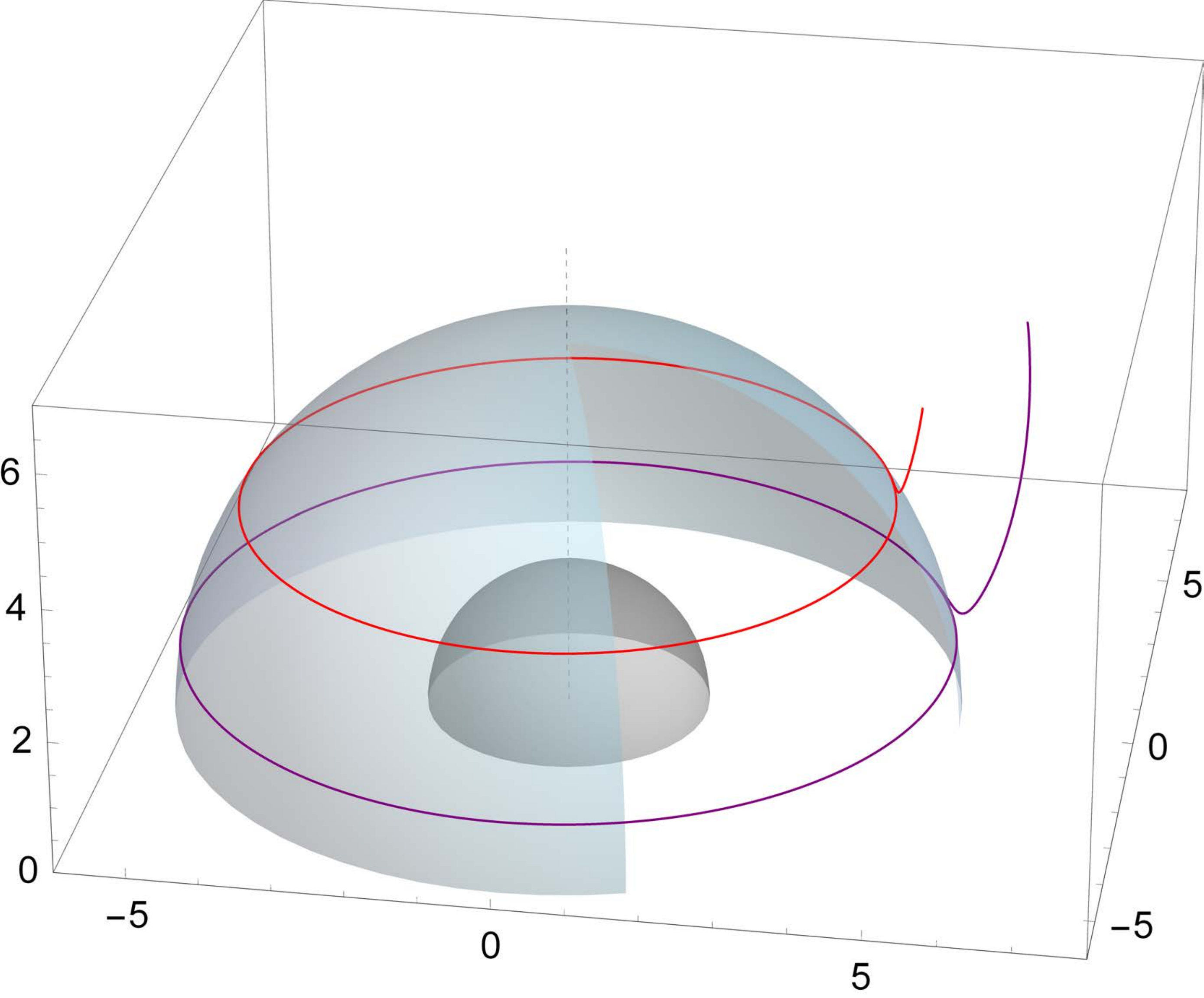}
\hspace{0.5cm}
\includegraphics[scale=0.3]{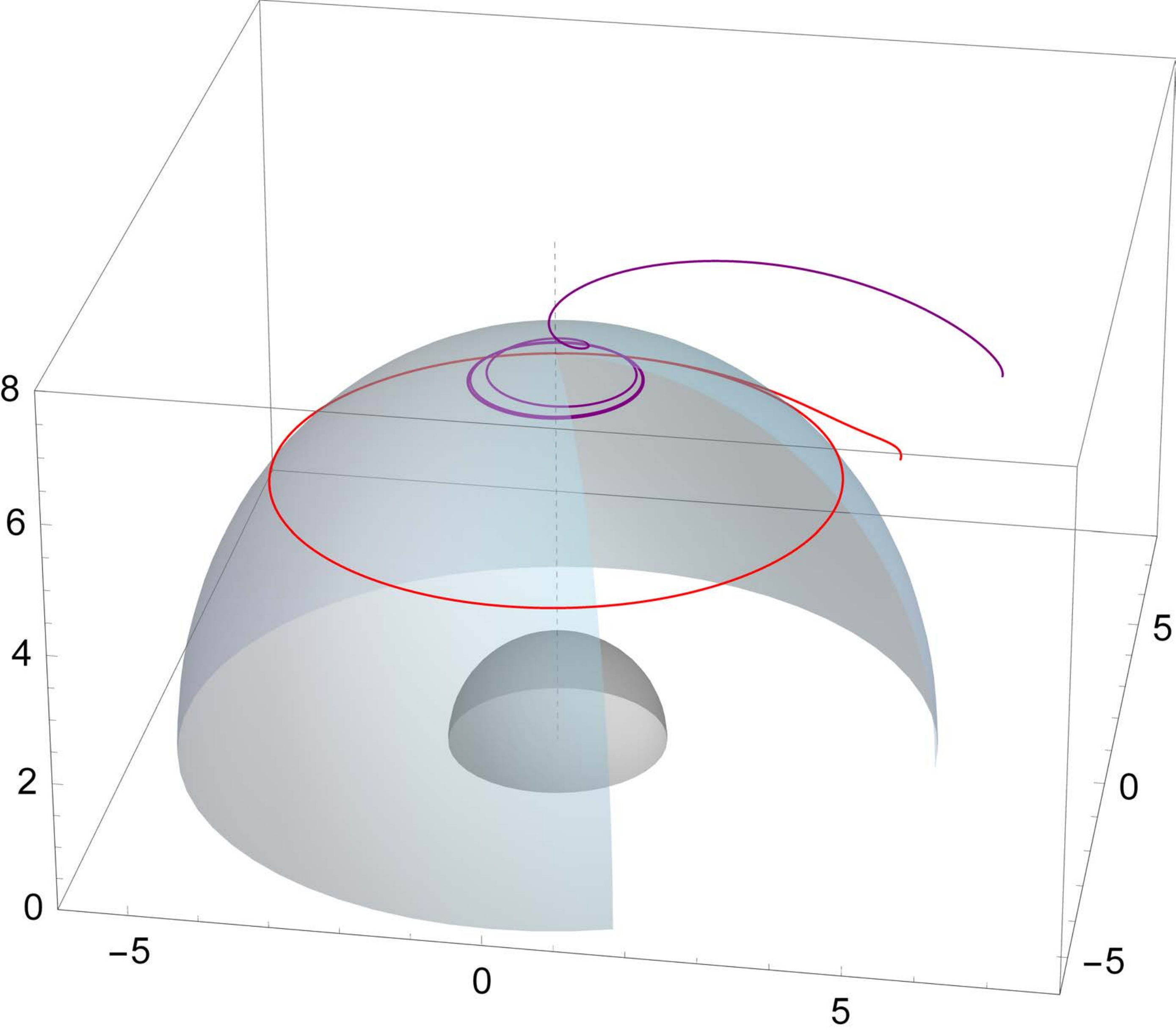}}
\caption{Off-equatorial circular orbits of test particles with zero angular momentum bound to the critical hypersurface. The red curves correspond to the case of test particle trajectory starting from the rest ($\nu=0, \gamma=1$) at $r_0=7M,\,\theta_0=\pi/4$ while the violet ones correspond to the case of test particle trajectory starting from the rest at $r_0=10M,\,\theta_0=\pi/4$. In the right panel the trajectories are plotted for the case of very small spin $a=0.05$ while in the left panel corresponds to the case of almost extreme spin $a=0.9995$. The relative luminosity of the radiating field takes the value of $A=0.8$. The inner black surface denotes the location of the north hemisphere of the event horizon. The blue-gray, partially open surface denotes the location of the north hemisphere of the critical hypersurface.}
\label{fig:Fig_traj_bounded}
\end{figure*}

\begin{figure*}[th!]
\centering
\hbox{
\includegraphics[scale=0.36]{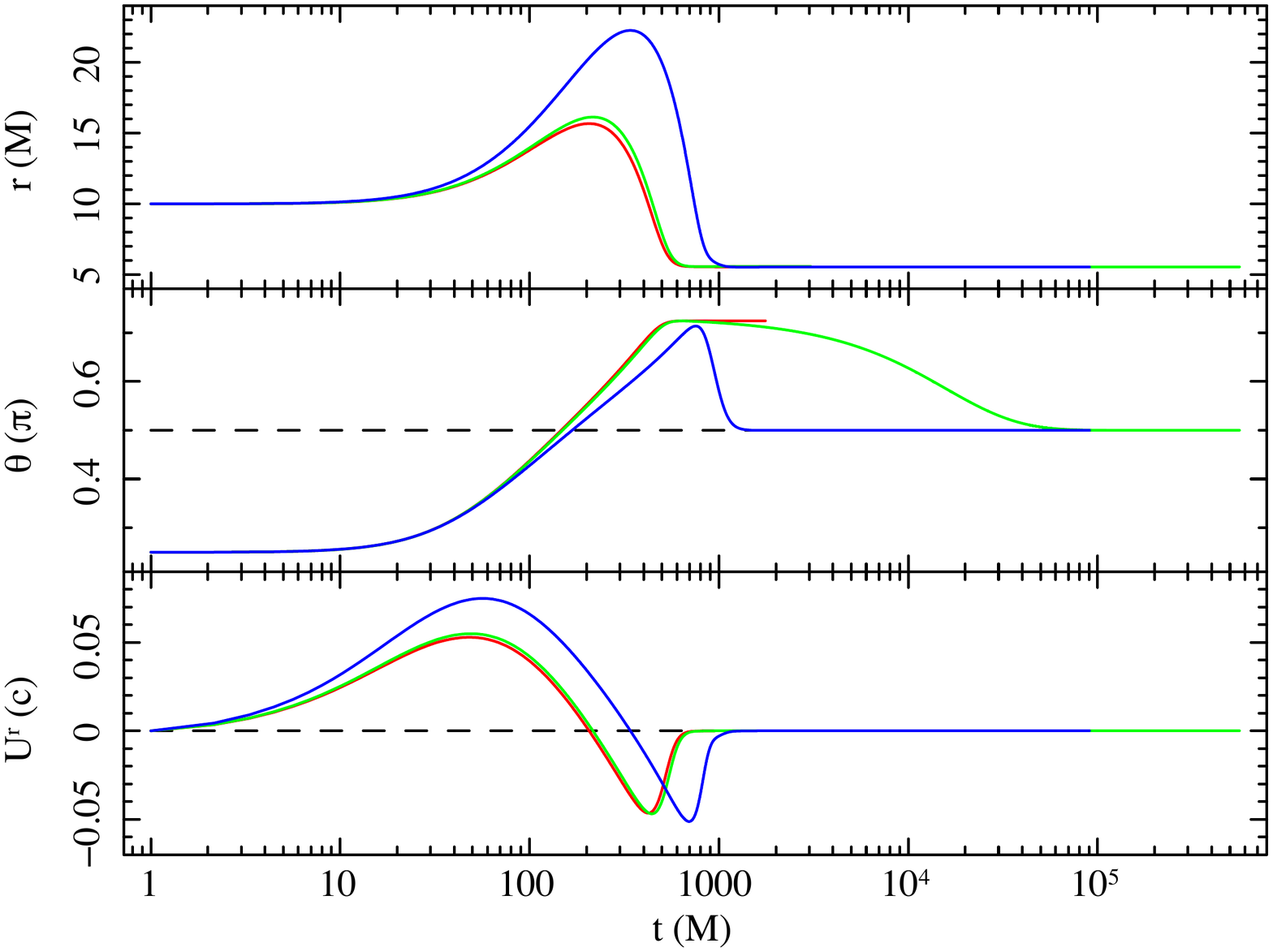}
\hspace{-1.5cm}
\includegraphics[scale=0.36]{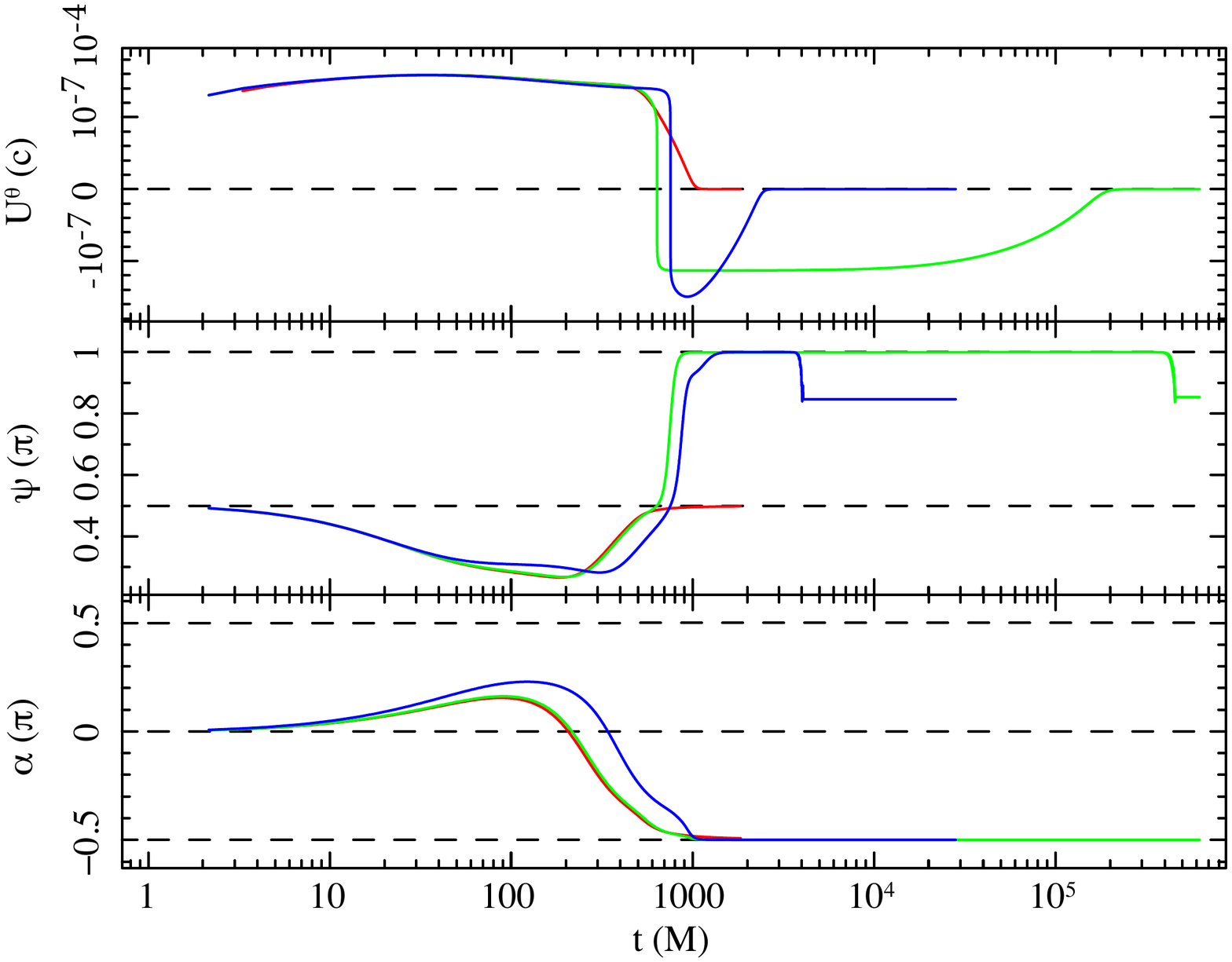}}
\caption{Profiles of $r$ and $\theta$ coordinates, $r$ and $\theta$ components of four-velocity ($U^r,\ U^{\theta}$), and $\psi$, $\alpha$ ZAMO local angles as functions of coordinate time $t$ for the test particles reaching the critical hypersurface with non-zero angular momentum ($\alpha \not=\pm\pi/2$). The function $u^\theta$ has been plotted in symmetric loagrithmic scale. The test particles are emitted outside of the critical hypersurface at $r_0=10M\,\theta_0=\pi/4$ in the azimuthal direction with the initial velocity $\nu_0=0.25$. The plots are constructed for the Schwarzschild case with zero spin case (red curves; compare to the left panel of Fig \ref{fig:Figs_S}), for the Kerr case with very small spin $a=0.05$ (green curves; compare to the right panel of Fig. \ref{fig:Figs_KSI1}) and for the Kerr case with almost extreme spin case $a=0.9995$ (blue curves; compare to the left panel of Fig \ref{fig:Figs_KH2}). The plots clearly illustrate the behavior in the touching point, where the test particles reaches the critical hypersurface. In the Kerr cases, $U^{\theta}$ is zero and ZAMO local polar angle $\psi$ takes the value of is $\pi/2$ at the touching point. Then during the latitudinal drift the angle $\psi$ increases while ZAMO local polar angle $\alpha$ decreases as the angular momentum of the test particle is removed. The local angle $\psi$ reaches the maximum value $\pi$. However, when the orbit is stabilized in equatorial plane and angular momentum is fully removed the angle $\psi$ is going back to the value of $\pi/2$ (The numerical integration of the trajectory is stopped earlier when spatial velocity is less than $10^{-20}$). In the Schwarzschild case, where latitudinal drift does not occur, the angular momentum is removed during the approaching to the critical hypersurface.}
\label{fig:Fig_plot}
\end{figure*}

In this section, we investigate in greater detail test particle trajectories bound to the critical hypersurface and their latitudinal drift towards the equatorial plane. We first emphasise that the condition for the radial balance of (outward) radiation force and gravitational attraction given by the equation (\ref{r-equilibrium}) is satisfied also when the test particle reaches the critical hypersurface with non-zero angular momentum (with its space velocity vector thus forming an arbitrary angle $\alpha$\ in the azimuthal direction; $\alpha \neq\pm\pi/2,\ \psi=\pm\pi/2,\ \nu=0,\ \gamma=1 $). The orbits of the test particles, reaching the critical hypersurface, can be divided into two classes with qualitatively different behavior.
\begin{itemize}
\item[(I)] Test particles with zero angular momentum achieve a complete balance of all forces acting at the critical hypersurface. Such case corresponds to  test particle trajectories which satisfy the condition $\nu=0$ at any $r(0)=r_0$ and $\theta(0)=\theta_0$ and are thus carried around by frame dragging in the azimuthal direction, along with  photons of the radiation field. At the critical hypersurface, such test particles then move along the purely off-equatorial circular orbits at constant latitude with angular velocity  $\Omega_{\mathrm{ZAMO}}$ (see Fig. \ref{fig:Fig_traj_bounded} and Fig. \ref{fig:Fig_plot}) remaining at rest relative to the appropriate ZAMO frame.
\item[(II)] Test particles that reach the critical hypersurface while still endowed with residual (non-zero) angular momentum (not coaligned with the spin axis, $\alpha \neq\pm\pi/2,\ \psi=\pm\pi/2, \theta\neq\pi/2$) attain radial balance, but  the PR effect still operates on them because the radiation field is  not yet directed in the radial direction in the test particle frame \cite{Bini2011}. Such particles exhibit a latitudinal drift on the critical hypersurface under the influence of the polar components of acceleration and consequently experience a polarly-oriented dissipative force originating from the interaction with the radiation field (see Eq. \ref{EoM2}). In the latitudinal drift the residual angular momentum of the test particle is progressively removed. Then in accordance with the reflection symmetry of the Kerr spacetime, full equilibrium ($\alpha=\psi=\pi/2, \nu=0, \gamma=1$) is attained in the equatorial plane, where latitudinal drift stops, the motion stabilises in a circular orbit and the angular momentum of the test particle is completely removed (see right panel of Fig. \ref{fig:Figs_S}, Figs \ref{fig:Figs_KSI1} - \ref{fig:Figs_KH2} and Fig. \ref{fig:Fig_plot}).
\end{itemize}
We note that in the Schwarzschild case, the spin and polar acceleration are absent, and thus latitudinal drift does not occur (see left panel of Fig. \ref{fig:Figs_S}).

\section{Conclusions}
\label{sec:end}
We developed a fully general relativistic treatment of the 3D PR effect in the Kerr geometry, therefore extending previous works describing 2D PR motion in the equatorial plane of relativistic compact objects. The outgoing radiation field we adopted assumes that photons propagate radially  with respect to the ZAMO frames. Such a boundary condition implies a purely radial propagation of the photons in any local ZAMO frame and may be considered as a simple approximation of the radiation field from a static emitting source very close to the horizon of a Kerr BH.
 
The resulting equations of motion for a test particle moving in the 3D space consist of a system of six coupled ordinary, highly nonlinear differential equations of first order. The non-linearity arises because of the general relativistic environment, further complicated by the PR effect which is a dissipative process and thus always entails nonlinearity. This set of equations is consistent with the previous 2D case for both test particles and photons moving in the equatorial plane \cite{Bini2009}.

Our analytical and numerical calculations in both the Schwarzschild and Kerr metric, show that 3D PR orbits are strongly affected by general relativistic effects, including frame-dragging. We have demonstrate the existence of a critical hypersurface, where the attraction of gravity is balanced by the outgoing radiation forces. In the case of the Schwarzschild geometry, the critical hypersurface is a sphere, on which the test particles are captured and remain at rest. In the case of the Kerr spacetime (with non-zero spin), the critical hypersurface is elongated in the polar direction. Test particles that are captured by it are dragged at an azimuthal angular velocity $\Omega_{\mathrm{ZAMO}}$, and, if still endowed with a residual and offset angular momentum, they exhibit a latitudinal drift that lead to spiraling towards the equatorial plane. Analysis of the $\nu$ profile shows that the test test particle spatial velocity attain that of the local ZAMO in an infinite time (see Figs. \ref{fig:Figs_KSI2} -- \ref{fig:Figs_KSO2} -- \ref{fig:Figs_KH1}). The test particle approaches the equatorial plane and radius of the hypersurface asymptotically. In future works we plan to relax some of the simple assumptions of the present study (e.g. by adopting more realistic radiation fields) and to investigate some possible astrophysical applications.  

\section*{Acknowledgements}
V.D.F. thanks the International Space Science Institute in Bern for support and the Max Planck Institute f\"ur Radioastronomie in Bonn for hospitality, since part of this work has been carried out there. V.D.F. thanks the Silesian University in Opava for partially funding this work. V.D.F. acknowledges useful discussions  with Dr. Andrea Geralico on the relativity of observer splitting formalism. V.D.F. is grateful to Prof. Antonio Romano for the valuable discussions. P.B. and D.L. acknowledge the Czech Science Foundation (GA\v{C}R) grant GA\v{C}R 17-16287S. P.B. and E.B. thank the International Space Science Institute in Bern for the hospitality to carry out part of this work. D.L. acknowledges SU SGS/15/2016 and MSK 03788/2017/RRC grants. L.S. acknowledges financial contributions from ASI-INAF agreements 2017-14-H.O and I/037/12/0 and from \emph{iPeska} research grant (P.I. Andrea Possenti) funded under the INAF call PRIN-SKA/CTA (resolution 70/2016).

\begin{appendix}
\section{Classical 3D Poynting-Robertson effect}
\label{sec:classicPR}
The classical radiation drag force was described and introduced by Poynting (1903) \cite{Poynting1903} and Robertson (1937) \cite{Robertson1937} in the 2D case. We extend the planar motion to the 3D case, written in spherical coordinates, $(r,\theta,\varphi)$. Noting that the classical drag force can be seen as a viscous effect depending linearly on the test particle velocity \cite{Poynting1903,Robertson1937,Defalco2018}, and assuming that the radiation 
propagates radially in the whole 3D space, the test particle equations of motion read
\begin{eqnarray}
\ddot{r}-r\dot{\varphi}^2\sin^2\theta-r\dot{\theta}^2+\frac{GM-Ac}{r^2}&=&-2A\frac{\dot{r}}{r^2},\label{eqm1}\\
r\ddot{\theta}+2\dot{r}\dot{\theta}-r\dot{\varphi}^2\sin\theta\cos\theta&=&-A\frac{\dot{\theta}}{r}\label{eqm2},\\
r\ddot{\varphi}\sin\theta+2\dot{r}\dot{\varphi}\sin\theta+2r\dot{\theta}\dot{\varphi}\cos\theta&=&-A\frac{\dot{\varphi}\sin\theta}{r},\label{eqm3}
\end{eqnarray}
where the dot means the derivative with respect to the time, $G$ is the gravitational constant, $M$ the mass of the central object, $c$ the speed of the light, and $A=Sd^2/(6c^2\rho a)$ is the luminosity parameter with $S$ being the surface luminosity density of the compact object and $d$ the distance Earth -- compact object. The term $-Ac/r^2$ represents the radiation pressure and $-2A\dot{r}/r^2$ is the specific angular momentum removed from the test particle due to the PR drag force.  

\subsection{Weak field approximation of the general relativistic equations}
We show here the way in which the 3D general relativistic equations of motion $ma(U)^\alpha=F_{\rm(rad)}(U)^\alpha$, Eqs. (\ref{EoM1}) -- (\ref{EoM6}), reduce to the classical 3D case, Eqs. (\ref{eqm1}) -- (\ref{eqm3}), in the weak field limit ($a\to0,\ r\to\infty,\ \nu/c\to0$). Eqs. (\ref{EoM4}) -- (\ref{EoM6}) are by definition
\begin{equation}
\begin{aligned}
&U^r\equiv\dot{r}\approx\nu\sin\psi\sin\alpha,\quad 
U^\theta\equiv \dot{\theta}\approx\frac{\nu\cos\psi}{r},\\
&\qquad U^\varphi\equiv \dot{\varphi}\approx\frac{\nu\sin\psi\cos\alpha}{r\sin\theta}. 
\end{aligned}
\end{equation}
The radial components of the ZAMO kinematical quantities reduce to
\begin{equation}
\begin{aligned}
&a(n)^{\hat r}\approx \frac{M}{r^2},\quad \theta(n)^{\hat r}{}_{\hat\varphi}\approx0,\\
&\qquad k_{\rm (Lie)}(n)^{\hat r}\approx -\frac{1}{r}, 
\end{aligned}
\end{equation}
expressed in geometrical units $G=c=1$, and where the relative Lie radial curvature reduces to the curvature of the osculating sphere (see \cite{Bini2009,Bini2011,Defalco2018}, for comparisons). Instead for the polar components of the ZAMO kinematical quantities we have 
\begin{equation}
\begin{aligned}
&a(n)^{\hat \theta}\approx 0,\qquad \theta(n)^{\hat r}{}_{\hat\varphi}\approx0,\\
& k_{\rm (Lie)}(n)^{\hat \theta}\approx -\frac{1}{r\tan\theta}, 
\end{aligned}
\end{equation}
where the relative Lie polar curvature describes the longitudinal Euler acceleration \cite{Bini1997a,Bini1997b,Defalco2018}. Now it is easy to see how the test particle acceleration, $a(U)^\alpha$, reduces to the left members of Eqs. (\ref{eqm1}) -- (\ref{eqm3}). Approximating the radiation force, $F_{\rm(rad)}(U)^\alpha$, through linear terms in the velocity field, we have (see \cite{Defalco2018}, for comparisons)
\begin{equation}
\begin{aligned}
&F_{\rm(rad)}(U)^{\hat r}\approx \frac{A}{r^2}(1-2\dot{r}),\\  
&F_{\rm(rad)}(U)^{\hat \theta}\approx-\frac{A}{r}\dot{\theta},\\
&F_{\rm(rad)}(U)^{\hat \varphi}\approx-\frac{A}{r}\dot{\varphi}\sin^2\theta,
\end{aligned}
\end{equation}
which reduce to the right members of Eqs. (\ref{eqm1}) -- (\ref{eqm3}). We note that the time component of the equations of motion, $ma(U)^t=F_{\rm(rad)}(U)^t$, reduces to \cite{Defalco2018}
\begin{equation} 
\frac{d}{dt}\left(\frac{\nu^2}{2}+\frac{A-M}{r}\right)=-A\frac{\nu}{r^2}-A\frac{\dot{r}^2}{r^2},
\end{equation}
which represents the energy conservation equation. Indeed, the left term represents the total mechanical energy, while the right term corresponds to the dissipated energy.
\end{appendix}

\bibliography{references}

\end{document}